\documentclass{aa}  

\usepackage{graphicx}
\usepackage{txfonts}
\begin{document}

\title
{
Interferometric observations of SiO thermal emission in the inner wind of M-type AGB stars
IK Tauri and IRC+10011\thanks{This work is based on observations carried out under project number W15BJ with the IRAM
NOEMA interferometer. IRAM is supported by INSU/CNRS (France), MPG (Germany), and IGN (Spain).
}\thanks{
Reduced datacubes are only available in electronic form at the CDS via anonymous ftp to cdsarc.u-strasbg.fr (130.79.128.5)
or via  http://cdsweb.u-strasbg.fr/cgi-bin/qcat?J/A+A/
}
}

   \author{J. L. Verbena
          \inst{1,2}
           \and
          V. Bujarrabal
          \inst{3}
          \and  
          J. Alcolea
         \inst{2}
        \and
         M. G\'omez-Garrido
        \inst{4,3}
        \and
        A. Castro-Carrizo
        \inst{5}    
          }

   \institute{
Molecular Astrophysics Group, Instituto de F\'isica Fundamental (IFF-CSIC), C/Serrano 123, 28006, Madrid, Spain
\\
\email{jl.verbena@csic.es}
\and
Observatorio Astron\'omico Nacional (OAN-IGN), C/Alfonso XII 3, 28014, Madrid, Spain
\and
Observatorio Astron\'omico Nacional (OAN-IGN), Apdo 112, 28803, Alcal\'a de Henares, Madrid, Spain 
\and    
Instituto Geogr\'afico Nacional, Centro de Desarrollos Tecnol\'ogicos, Observatorio de Yebes, Apdo 148, 19080 Yebes, Spain
\and
Institut de Radioastronomie Millim\'etrique, 300 rue de la Piscine, 38406, Saint Martin d'H\`eres, France
}

   \date{Received; accepted}

\titlerunning{Interferometric Observations of IK Tau and IRC+10011}

 
  \abstract
{
Asymptotic giant branch (AGB) stars go through a process of strong mass loss that involves
pulsations of the atmosphere, which extends to a region in which the conditions are adequate 
for dust grains to form. Radiation
pressure acts on these grains which, coupled to the gas, drive a massive outflow. The details of this process
are not clear, including which molecules are involved in the condensation of dust grains.  
}
   {
We seek to study the role of the SiO molecule in the process of dust formation
and mass loss in M-type AGB stars.
}
{
Using the IRAM NOEMA interferometer we observed the $^{28}$SiO and $^{29}$SiO $J=3-2$, $v=0$ emission from
the inner circumstellar envelope of the evolved stars IK Tau and IRC+10011.
We computed azimuthally averaged emission profiles to
compare the observations to models using a molecular excitation and ray-tracing code for SiO
thermal emission.
}
   {
We observe circular symmetry
in the emission distribution.
We also find that the source diameter varies only marginally 
with radial velocity, which is not the expected behaviour for
envelopes expanding at an almost constant velocity.
The adopted density, velocity, and abundance laws, together with the mass-loss rate, which best fit the observations,
give us information concerning the chemical behaviour of the SiO molecule and its role
in the dust formation process. 
}
   {
The results indicate that there is a strong coupling between the depletion of gas-phase SiO and
gas acceleration in the inner envelope. This could be explained by the condensation of SiO into
dust grains. 
}


   \maketitle
%

\section{Introduction}

During the asymptotic giant branch (AGB) phase, stars go through a process of
strong mass loss that is caused by the pulsating levitation
of the atmosphere, transporting gas to a region in which the conditions are right for dust condensation
(i.e. low temperature and high density). Dust
is then
accelerated by radiation pressure from the central star,
and as a consequence of the efficient coupling between gas and dust, drives a massive outflow. This outflow
becomes part of the interstellar medium after forming a
circumstellar envelope and a planetary nebula in future post-AGB phases \citep{Hofner2018}.
In a mature galaxy like the Milky Way, stars of low to intermediate mass
(those with an initial mass between 0.8 and 8 M$_{\odot}$) are responsible
for a considerable fraction of the reinsertion of material into the
galaxy due to such massive molecule-rich winds \citep{Busso1999}. 

AGB stars consist of a carbon-oxygen core, a helium-burning shell,
and a more external hydrogen-burning shell.\footnote{At the high mass end of AGB stars the core
may consist of oxygen and neon, the super AGB stars.}  
Eventually the helium shell runs out of fuel, but in periods of $\sim$ 10$^4$ to 10$^5$ years
the hydrogen burning shell provides the inner shell with enough helium
for an explosive event to take place, where
helium burns again. This causes the hydrogen shell to cease
burning and results in an expansion of the stellar atmosphere. These
periodic events are known as thermal pulses. In this way the star enters the thermally pulsing asymptotic giant branch (TP-AGB),
and the related third dredge-up events
lifts up carbon and other elements to the outer atmosphere.
Depending on the duration of the TP-AGB, which depends on the initial mass, 
some stars become C-rich ([C] > [O])
and some remain O-rich (i.e. M-type) during the whole evolution.
C-rich chemistry
is fairly well understood and has been shown to be feasible in stellar outflow conditions \citep{Gail1984}. 
But the same cannot be said about O-rich chemistry, 
for which the process of nucleation and growth of dust grains is not clear
(see e.g. \citealp{Gobrecht2016,Gail2016}). According to observations,
the dominating dust components formed in the photospheres of M-type stars are 
magnesium iron silicates
\citep{Molster2010}, and laboratory investigations
support this assessment \citep{Vollmer2009, Nguyen2010, Bose2010}.  
More recently it has been found that for objects with low mass-loss rates alumina dust may 
be as abundant as silicate dust \citep{Zhao2012, Karavikova2013, Khouri2015}.

Among all the gas-phase species of refractory elements that could be
involved in the dust nucleation process, SiO is one of the most abundant.  
Dust containing Si and O are also identified 
through the 9 ~$\mu$m and 18~$\mu$m silicate features in infrared spectra of stars
\citep{Forrest1975, Pegourie1985}.   
Therefore, the SiO gas-phase abundance should fall off as the molecules are incorporated into
dust grains. At the same time, the expansion velocity of the gas should increase as the
radiation pressure acts more efficiently on dust grains of larger size. 
By the determination of the abundance of SiO 
and its dynamical behaviour throughout the envelope
we can assess if a relation between SiO depletion and
dust formation exists. Other alternatives to explain dust formation in O-rich stellar atmospheres
have been studied recently by, for example
\citet{Goumans2012}, \citet{Plane2013}, and \citet{Bromley2014}. 

Thermal\footnote{We use the term ``thermal" to differenciate between $v=0$ emission and maser emission from 
vibrationally excited states ($v>0$), even though SiO excitation is normally far from
thermal equilibrium with the gas kinetic temperature so that radiative excitation plays
a non-negligible role} 
emission from the SiO molecule was previously studied by \cite{Bujarrabal1989} with
the Institut de Radioastronomie Millim\'etrique (IRAM) 30~m radio antenna and has given us information about the physical 
conditions of the circumstellar envelope.
These authors also observed the CO molecule and simultaneously
modelled the emission from both CO and SiO.
There are also interferometric observations of SiO that suggest that the
gas might not reach its final expansion velocity as fast as was previously believed
and that dust formation might also not occur as instantaneously \citep{Lucas1992}. 
These authors found a relatively large initial abundance of gaseous SiO,
$X$(SiO) $\sim$ $5 \times 10^{-5}$
, which has a strong decrease beyond $\sim$ $10^{15}$ cm from grain formation, which would itself be
associated with the gas acceleration\footnote{ 
We refer to the relative abundance of SiO with respect to H$_2$ as $X$(SiO).} (see also \citealp{Bujarrabal1991}).

\cite{Gonzalez2003} carried out a statistical analysis on a sample of AGB stars. They
found a relation between the radius of the SiO envelope and what they called the ``density measure'', 
$\dot{M}/V_e$, where $\dot{M}$ is the stellar mass-loss rate and $V_e$ is the expansion velocity. 
They deduced, in most objects, relatively lower inner abundances: $X$(SiO)~$\leq$~$10^{-5}$, 
and stressed the importance of photodissociation in establishing the outer SiO-rich boundary. On the other hand
\cite{Schoier2004} concluded that $X$(SiO) is as high as $4~\times~10^{-5}$ in the inner envelope and has a decrease that is higher than a factor 10, due to grain formation, at $\sim$~$10^{15}$ cm.
According to these authors, a small fraction of SiO remained in the gas until photodissociation.
\cite{Decin2010}  also modelled several emission lines from observations carried out with various radio telescopes
and obtained their own set of physical parameters for the envelope.
Later on \cite{Decin2018} carried out a spectral line and imaging survey to observe 
molecular species key to the formation of dust grains, providing constraints on the
properties of oxygen-dominated stellar winds. These authors concluded that properties tend
to be different for stars with high mass-loss rates as compared to those with low mass-loss rates and found a similar stratification of the SiO abundance starting with a somewhat lower
value.

We conducted observations of two evolved, O-rich, stars with intermediate mass-loss rates, 
IK Tau and IRC+10011 (a.k.a. WX Psc),
with 
the NOrthern Extended Millimeter Array (NOEMA) interferometer,
resulting in
higher spatial resolution than previously available. We describe the observations
and data reduction in Section \ref{sec:obse} and
discuss the results in Section \ref{sec:res}.   
We obtained the best fit for the
azimuthally averaged emission 
using a molecular excitation and line transfer code. 
The resulting models and physical parameters obtained are shown 
and discussed in Section \ref{sec:ana}, where we also discuss the implications for dust formation.
Final discussion and conclusions are shown in Sections \ref{sec:discu} and \ref{sec:conc}.

\section{Observations} \label{sec:obse}

Observations were carried out towards IK Tau and IRC+10011 with the NOEMA millimeter interferometer
(project W15BJ) during the winter and autumn of 2016 using seven and eight
antennas. The observations were performed by
alternating between the two sources
in cycles of $\sim$ 15 minutes in the so-called track-sharing mode.
Acquisitions were obtained from February 21 to 24, 2016 for the most
extended A configuration, and between October 10 to 31, 2016 for the C
configuration. In the merged data we obtained baselines ranging from 48 to
720~m, for a total of about 5 hours on each source. Data were recorded
with the Widex Correlator for 4~GHz bandwidth with
two polarisations and a channel spacing of 2~MHz.  In addition,
observations of one receiver polarisation were simultaneously processed
by the narrow-band correlator to achieve a spectral resolution of
0.3~MHz in a window of 110~MHz around the frequencies 130.269 and 128.637~GHz, which are the rest frequencies of the $J=3-2$, $v=0$ emission lines for 
$^{28}$SiO and $^{29}$SiO, respectively; these comprise 
our two lines of interest.

The data calibration was performed with CLIC, which
is part of the GILDAS\footnote{http://www.iram.fr/IRAMFR/GILDAS} package. 
Two phase calibrators were observed, each one
close to one of the sources, also in cycles of 15 minutes.  The bright
quasars 3C84 and 3C454.3 (with more than 15 Jy in continuum emission)
were observed for the bandpass calibration, and acquisitions in MWC349
and LKHA101 (the standard flux calibrators for the NOEMA observatory) were
used to reach an absolute flux accuracy in the calibrated data better
than 10\%. The achieved relative flux accuracy among our observed objects and transitions
is nonetheless better than 10\%.

\begin{figure}
\begin{center}
\includegraphics[trim=0 0 205 0, clip, width=.45\textwidth]{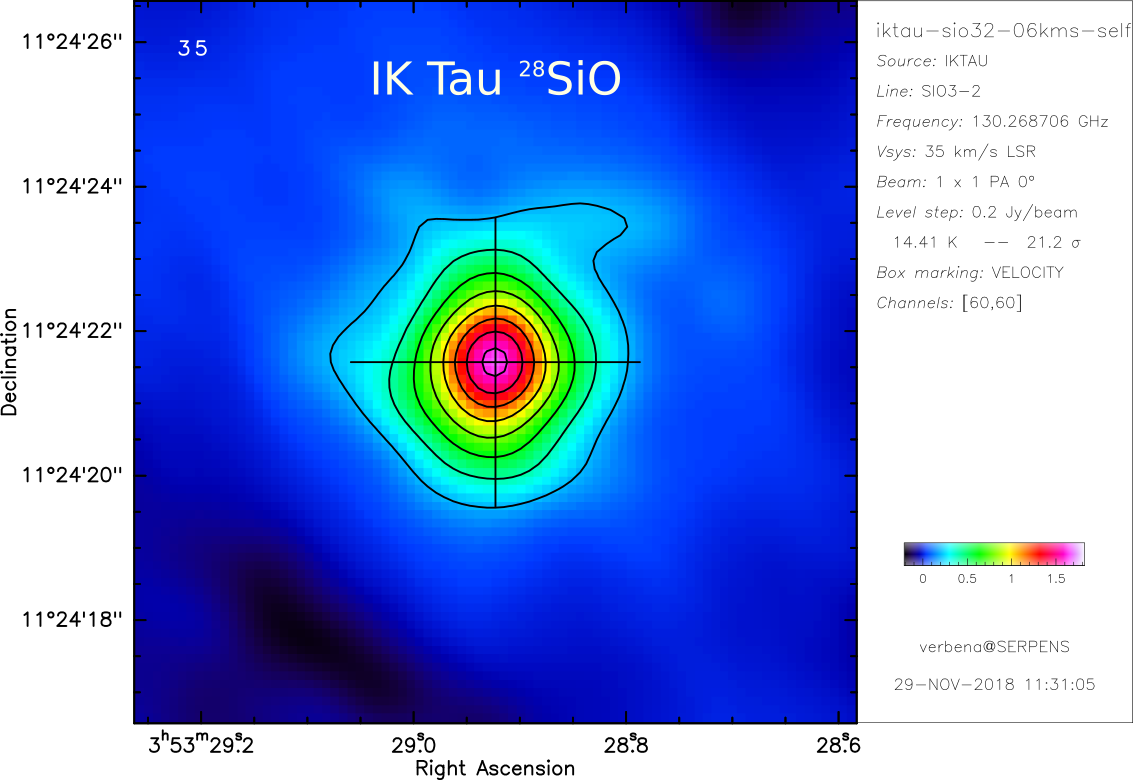}
\includegraphics[trim=0 0 205 0, clip, width=.45\textwidth]{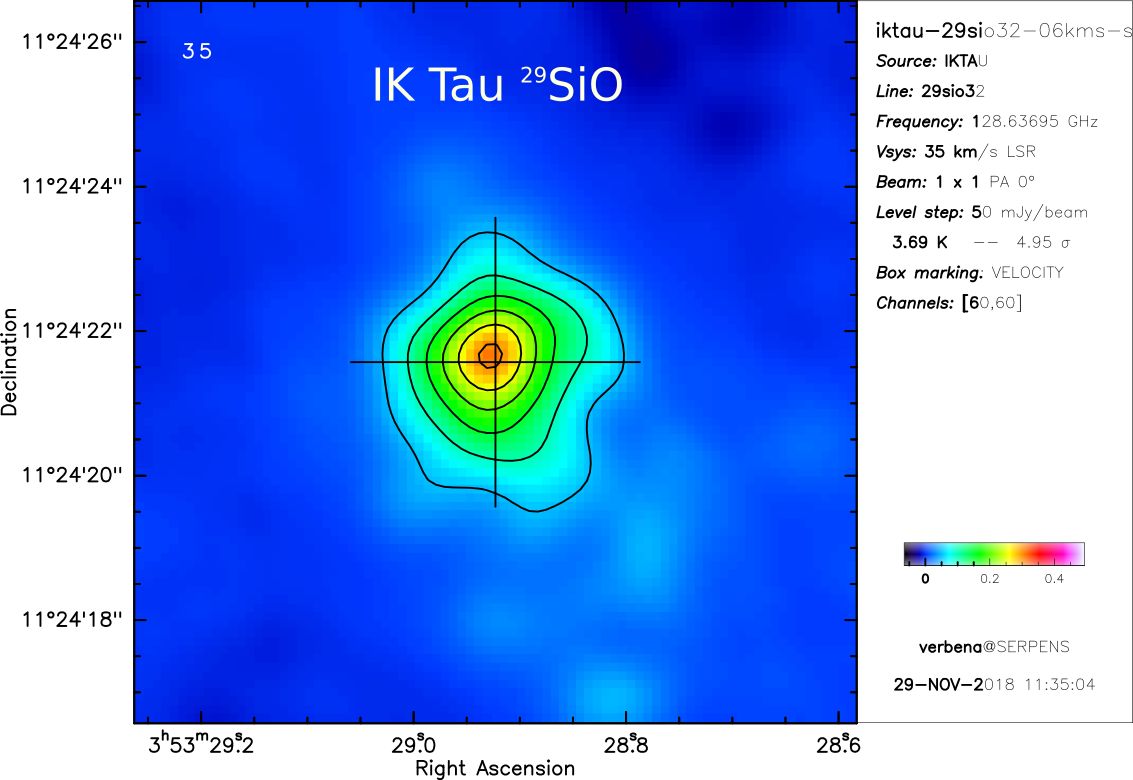}
\end{center}
\caption{
SiO emission maps in one of the central channels for IK Tau in $^{28}$SiO (top) and $^{29}$SiO (bottom). 
Contour level spacings are 200~mJy~beam$^{-1}$ for $^{28}$SiO and 50 mJy beam$^{-1}$ for $^{29}$SiO. 
In the top left corner the channel LSR velocity is indicated.
}
  \label{fig:observ}
\end{figure}

\begin{figure}
\begin{center}
\includegraphics[trim=0 0 205 0, clip, width=.45\textwidth]{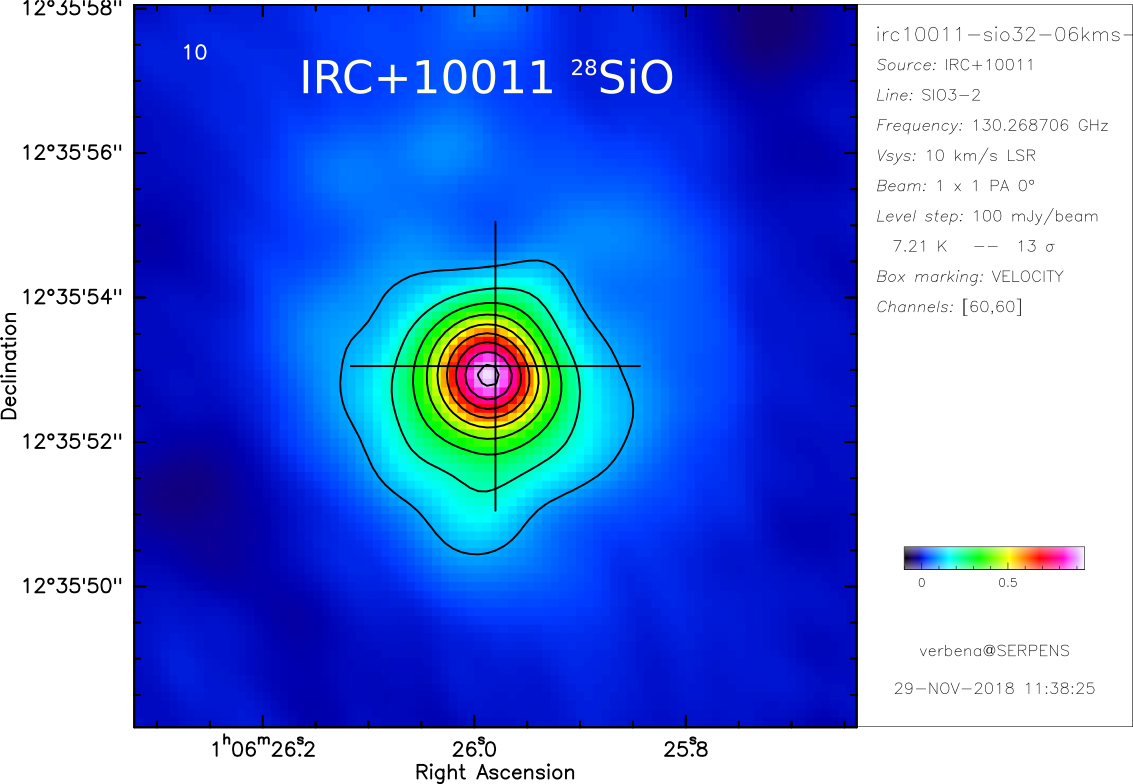}
\includegraphics[trim=0 0 205 0, clip, width=.45\textwidth]{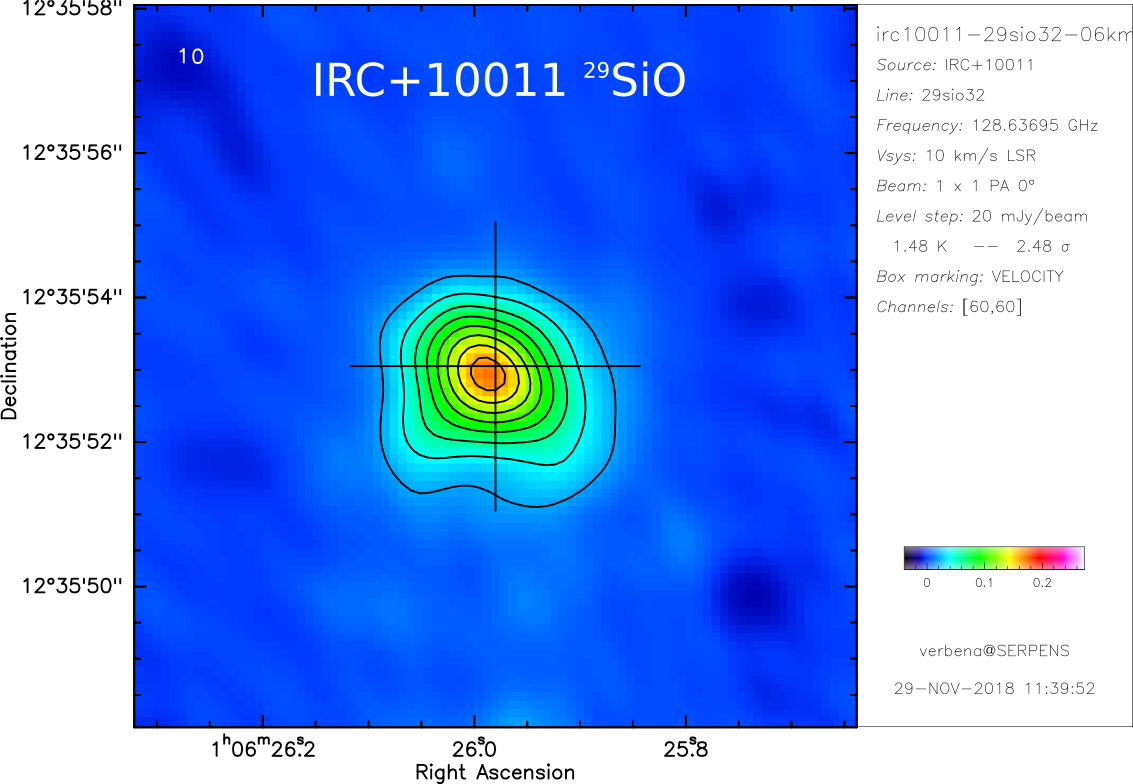}
\end{center}
\caption{
SiO emission maps in one of the central channels for IRC+10011 in $^{28}$SiO (top) and $^{29}$SiO (bottom). 
Contour level spacings are 100~mJy~beam$^{-1}$ for $^{28}$SiO and 20 mJy beam$^{-1}$ for $^{29}$SiO.
In the top left corner the channel LSR velocity is indicated.
}
  \label{fig:observ2}
\end{figure}

Maps in a central channel for both objects and both isotopologues are shown in
Figs. \ref{fig:observ} and \ref{fig:observ2}.
The continuum emission
was subtracted to better identify the weak emission at extreme velocities.
A synthetic half-power beam size of 1" $\times$ 1" in diameter was obtained with data robust weighting.
The detailed data inspection was performed in 120 channels with spectral resolutions of
0.6 km~s$^{-1}$.
Data reduction was carried out with standard reduction techniques
using MAPPING (also included in the GILDAS package).
The reported systemic velocities of the sources are 34.4 and 9.6 km~s$^{-1}$ for IK Tau and IRC+10011, 
respectively \citep{Gonzalez2003}.

\section{Results: Size and flux} \label{sec:res}

\begin{figure*}
\begin{center}
\includegraphics[width=.48\textwidth]{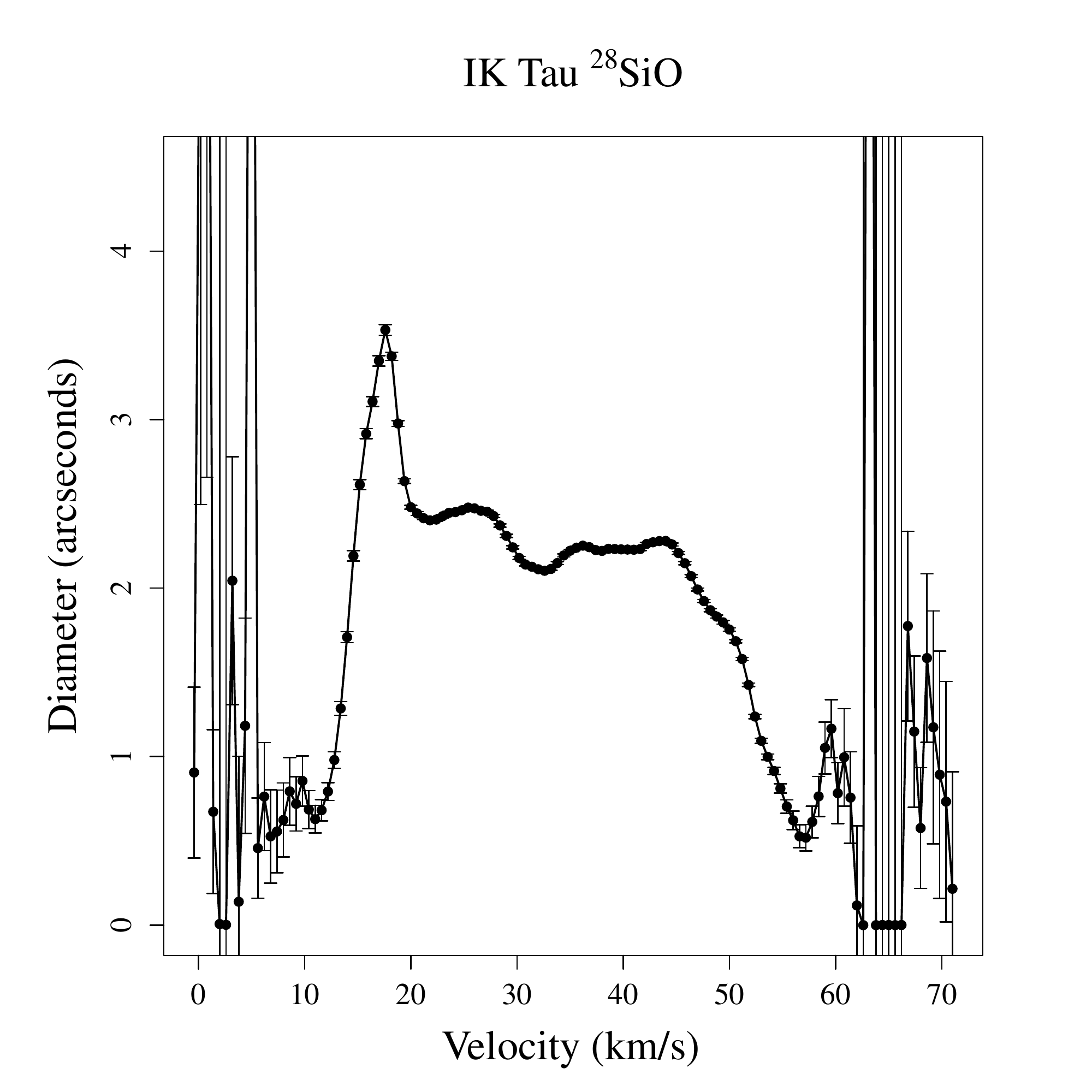}
\includegraphics[width=.48\textwidth]{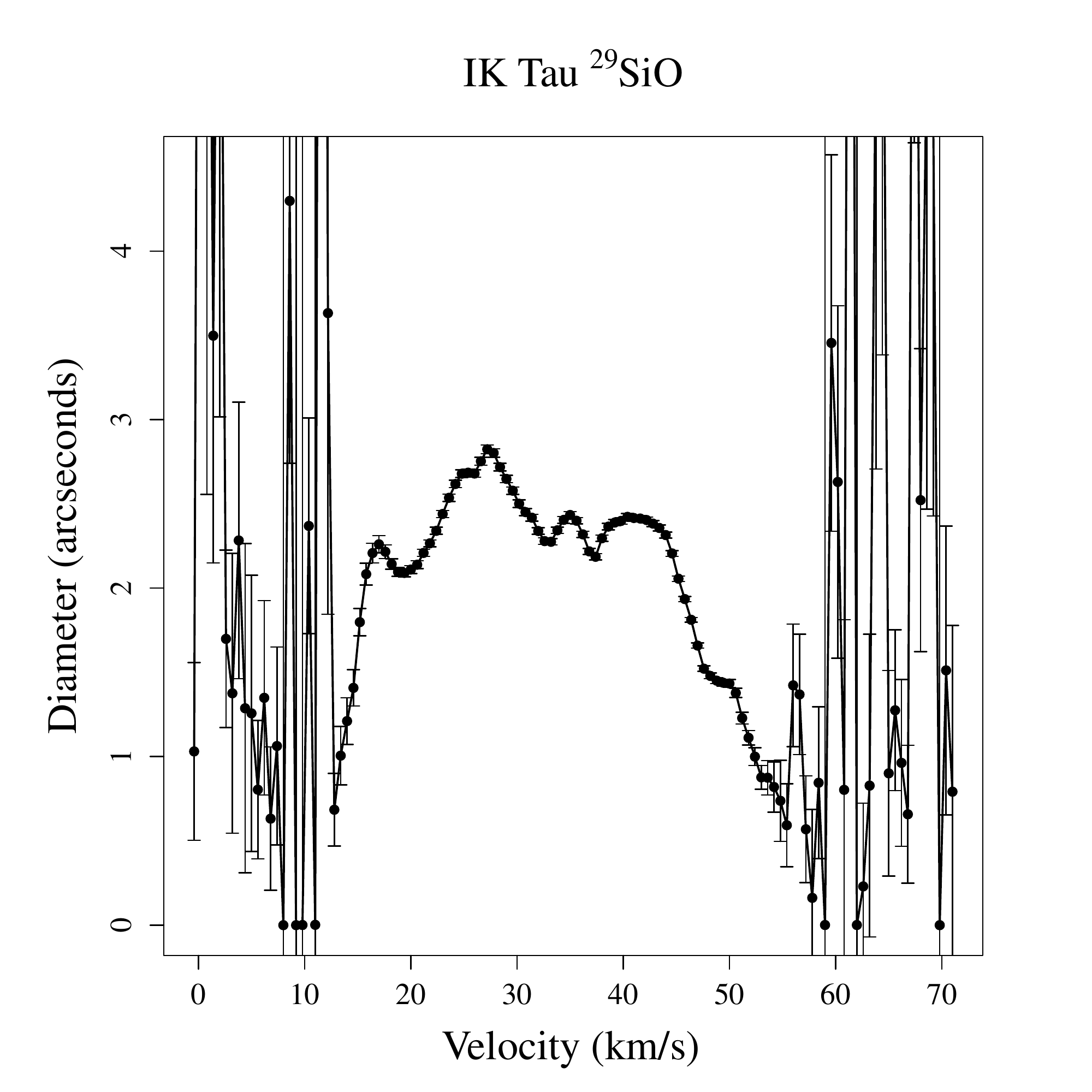}
\includegraphics[width=.48\textwidth]{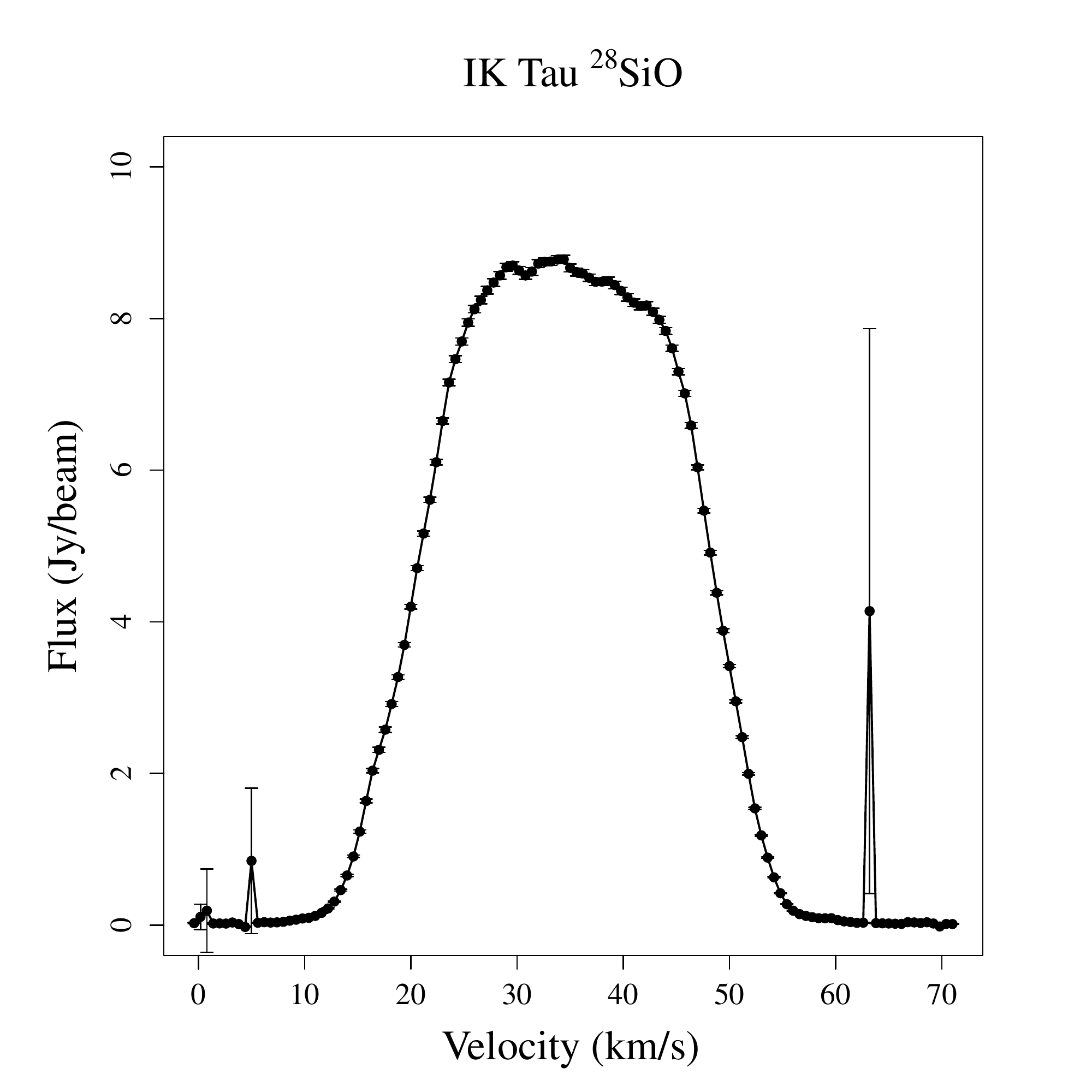}
\includegraphics[width=.48\textwidth]{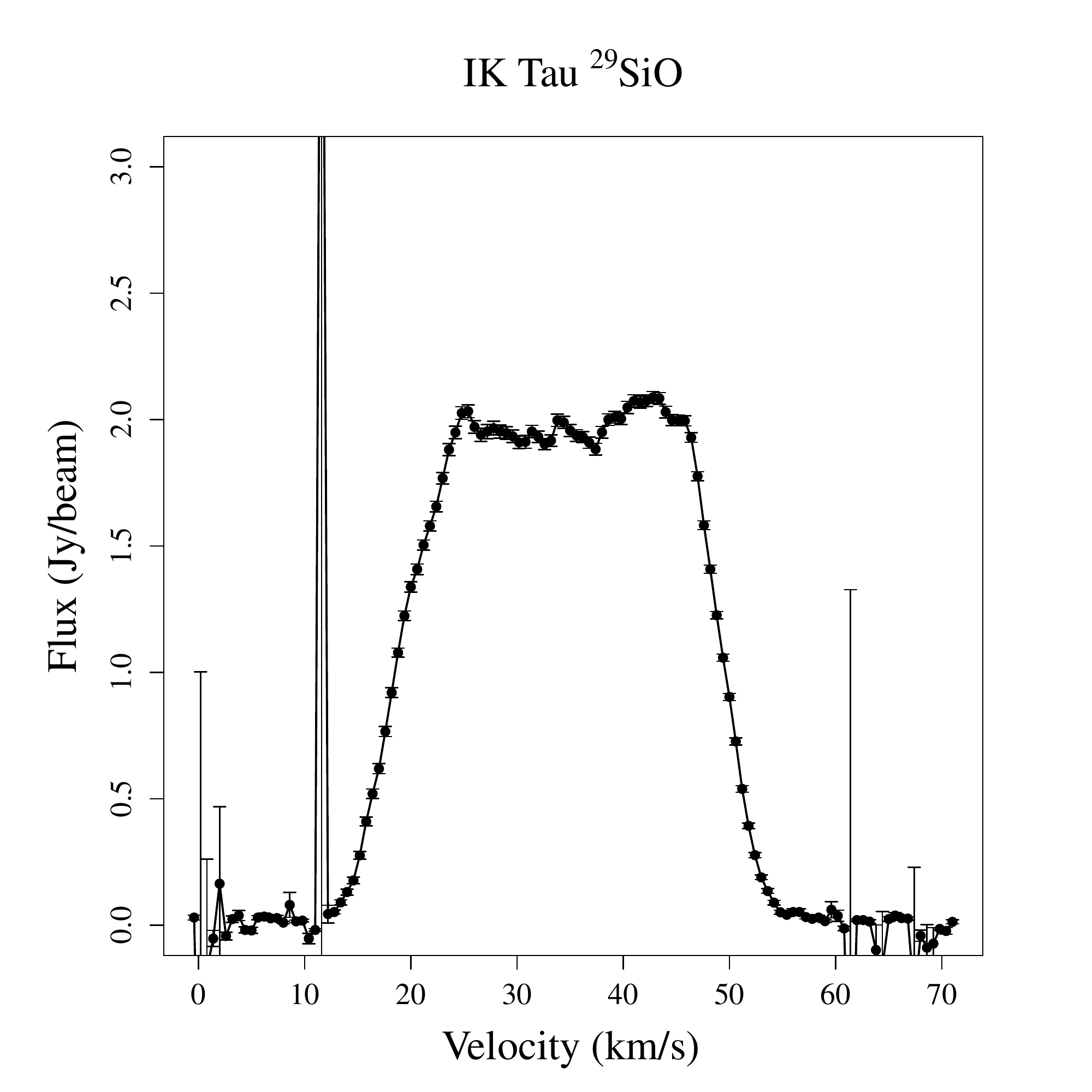}
\end{center}
\caption{
Half-power diameter and flux as functions of LSR velocity as determined from model fits
to the visibilities for IK Tau.  
As argued in \cite{Lucas1992}, the small variation of the diameter
as a function of velocity could be explained by a wind for which the final
expansion velocity is not yet fully reached.
}
  \label{fig:sizefluxik}
\end{figure*}

\begin{figure*}
\begin{center}
\includegraphics[width=.48\textwidth]{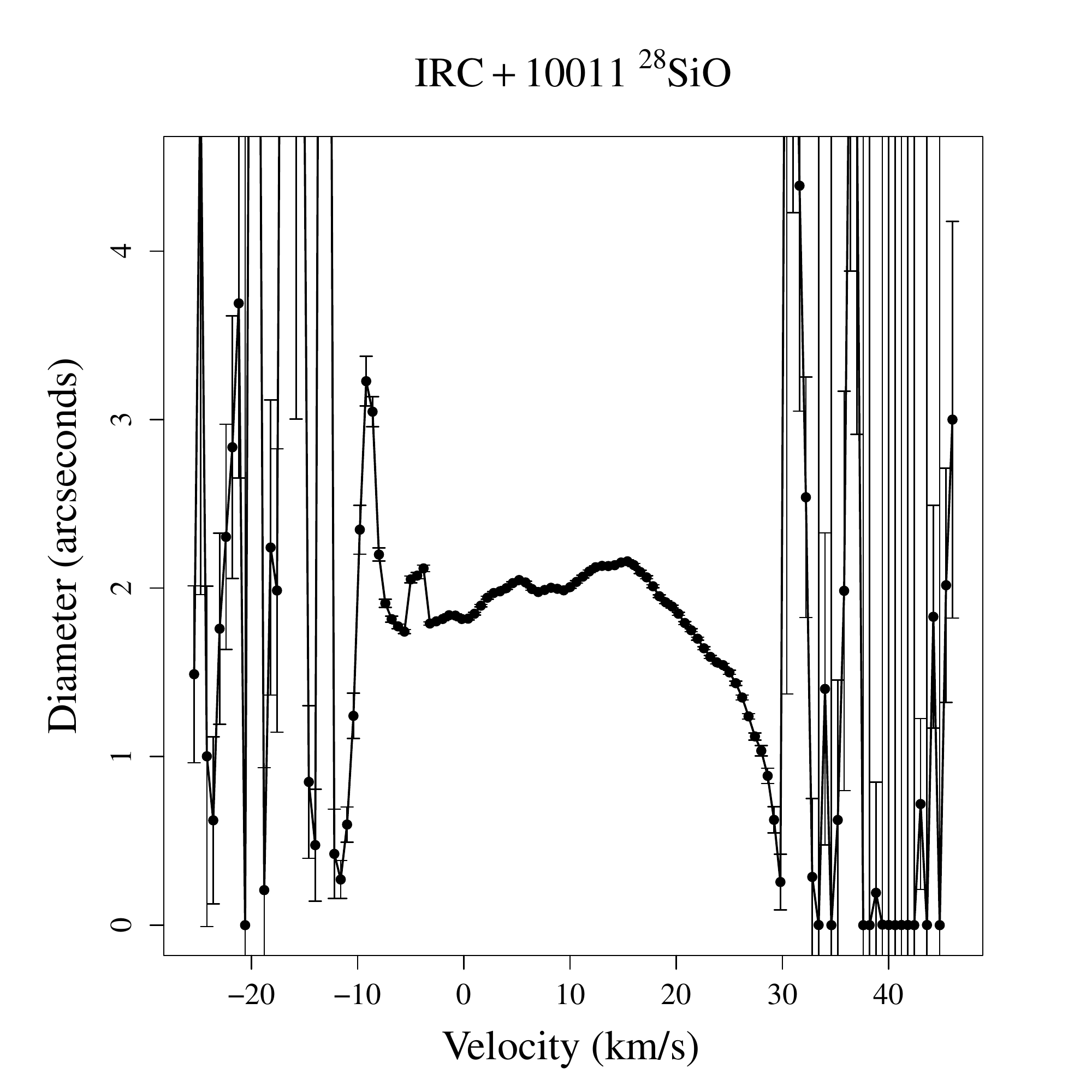}
\includegraphics[width=.48\textwidth]{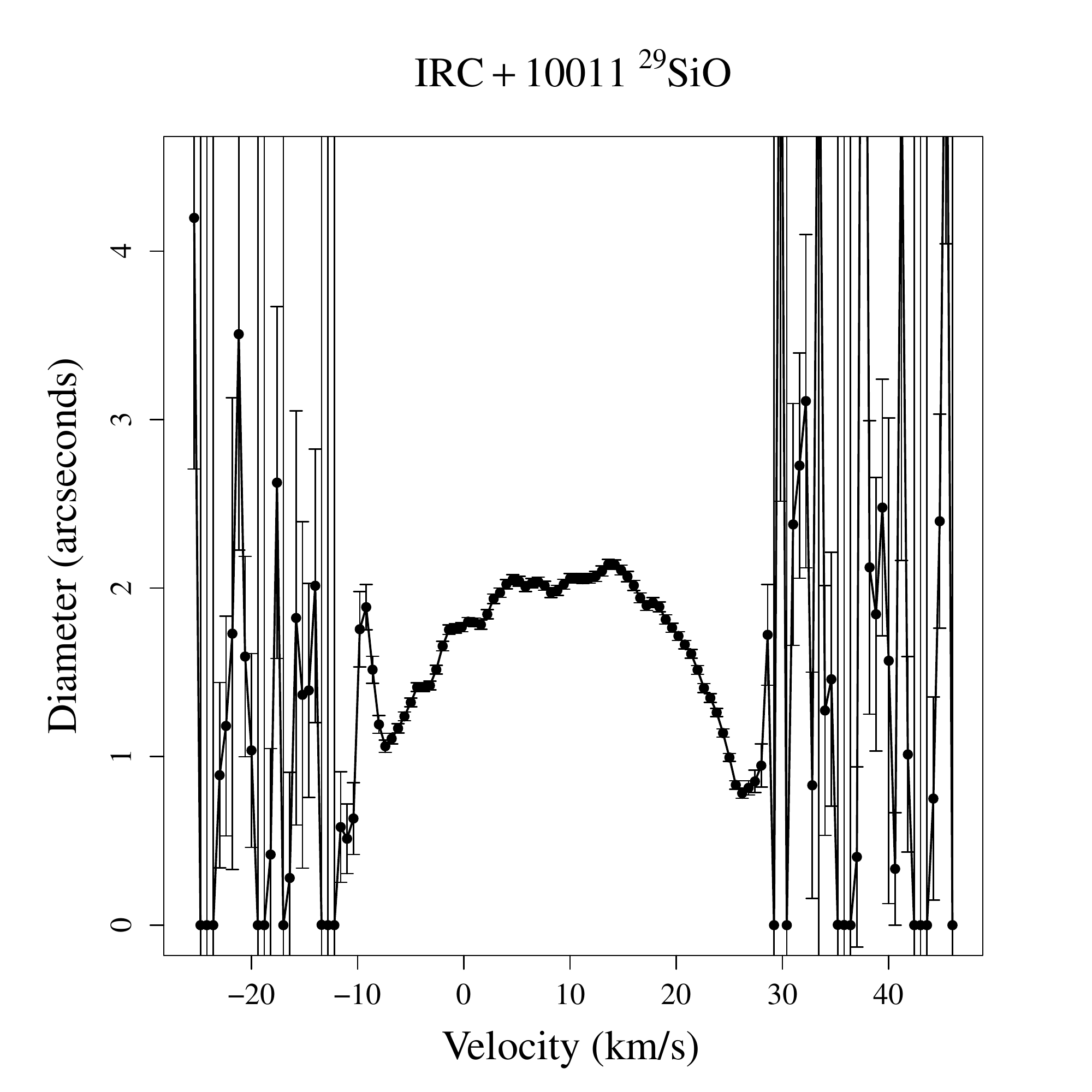}
\includegraphics[width=.48\textwidth]{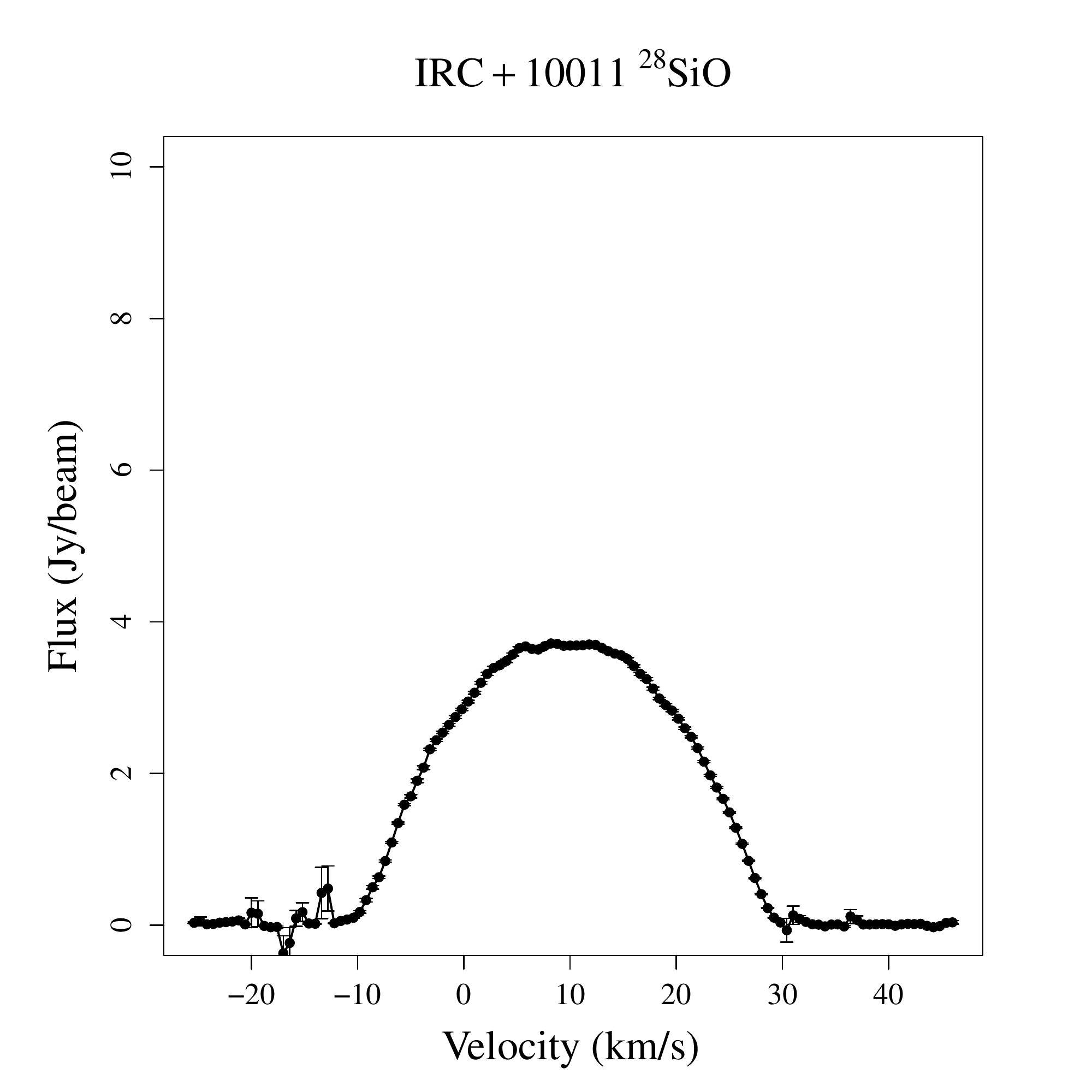}
\includegraphics[width=.48\textwidth]{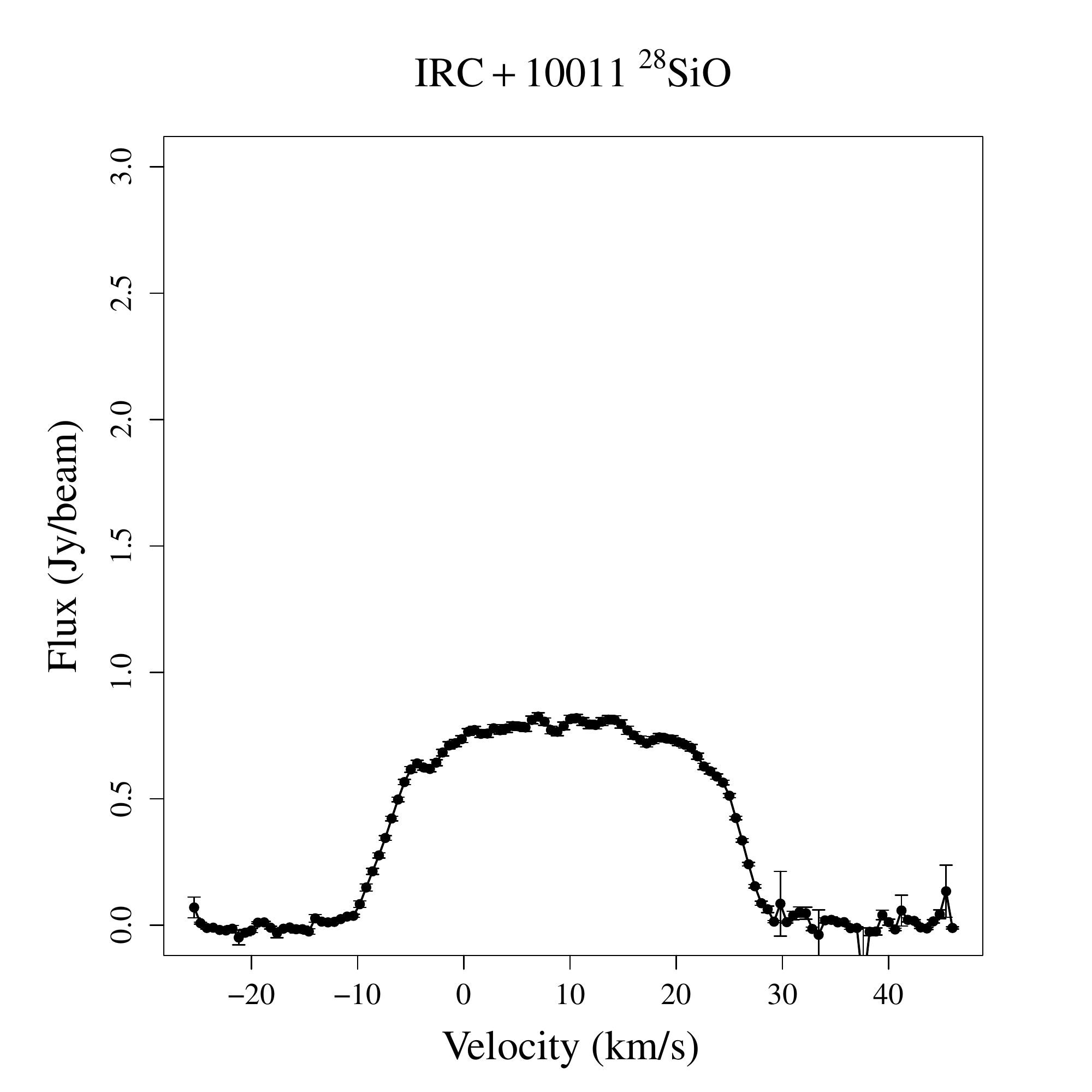}
\end{center}
\caption{
Half-power diameter and flux as functions of LSR velocity 
as determined from model fits to the visibilities for IRC+10011. 
The small variation in the diameter
as a function of velocity could be explained by a wind for which the final
expansion velocity is not yet fully reached.
}
  \label{fig:sizefluxirc}
\end{figure*}

In Figs. \ref{fig:sizefluxik} and \ref{fig:sizefluxirc}
we show the half-power angular diameters and fluxes as functions of local standard of rest (LSR) velocity determined 
from model fits to the visibilities obtained for IK Tau and IRC+10011.
The model used to represent the emission region was that of a Gaussian disc.
We note that the size of the 
SiO emission region is extended for
both objects, with angular diameters of $\sim$~2".5 for IK Tau and $\sim$~2" for IRC+10011, which translate
to physical diameters 
of the order of 1 $\times$ 10$^{16}$ cm and 2~$\times$~10$^{16}$~cm,
respectively, when considering the distances to the objects shown in Table \ref{tab:fixedparams}. 
  
The plots for both $^{28}$SiO and $^{29}$SiO show that the diameter does not decrease with
offset from the central velocity; instead, the diameter is very similar in
the line centres and wings. \cite{Lucas1992} argued that this can be explained if the final
expansion velocity has not been fully reached in the inner SiO emission region. In such a case the
hot, central regions emit only at the central velocities, while only the outer regions
are seen at the profile edges with a larger apparent size than if the velocity were constant.
On the contrary, \cite{Sahai1993} found no need for such slowly varying velocity fields. Instead,
they argued that this behaviour is a characteristic feature of power-law intensity distributions,
being scale-free, rather than Gaussian, where there are well-defined scale sizes.
Nonetheless, attempting other types of fits, such as uniform disc, exponential and power-law,
resulted in considerably larger errors in the fits to the observational data; therefore we 
believe that a Gaussian disc 
accurately represents the emission region and that indeed gas is still accelerating
in the inner envelope. 

\section{Radiative transfer and S\MakeLowercase{i}O abundance} \label{sec:ana}

To estimate the physical parameters that reproduce the emission we
used a radiative
transfer code
assuming spherical symmetry.
The parameters are fed into a molecular excitation and ray-tracing code that 
solves the statistical equilibrium equations by assuming the Sobolev approximation,
i.e. describing photon trapping by means of an escape-probability algorithm.
Level populations were then used to calculate the expected intensity for a high number
of rays in the direction of the line of sight, following the standard radiative
transfer equation. We take into account 40 rotational levels in the v = 0 vibrational state.  
The black-body stellar radiation field with stellar temperatures listed
in Table \ref{tab:freeparams} is taken into account for radiative excitation. 
The dust radiation field 
is not taken into account.
Finally, the brightness distributions are convolved with
the telescope beam to simulate the observational maps. Similar codes were used by
\citet{Bujarrabal1989} and \citet{Bujarrabal1991}, where more details on the
numerical treatment can be found. Collisional transition probabilities were
taken from the Leiden Atomic and Molecular Database
(LAMDA\footnote{http://home.strw.leidenuniv.nl/$\sim$moldata/})
\citep{Schoier2005}. 
We find that the level populations
are practically thermalised for $^{28}$SiO, but not for $^{29}$SiO owing to
its lower opacities.

\begin{figure*}
\begin{center}
\includegraphics[width=.48\textwidth]{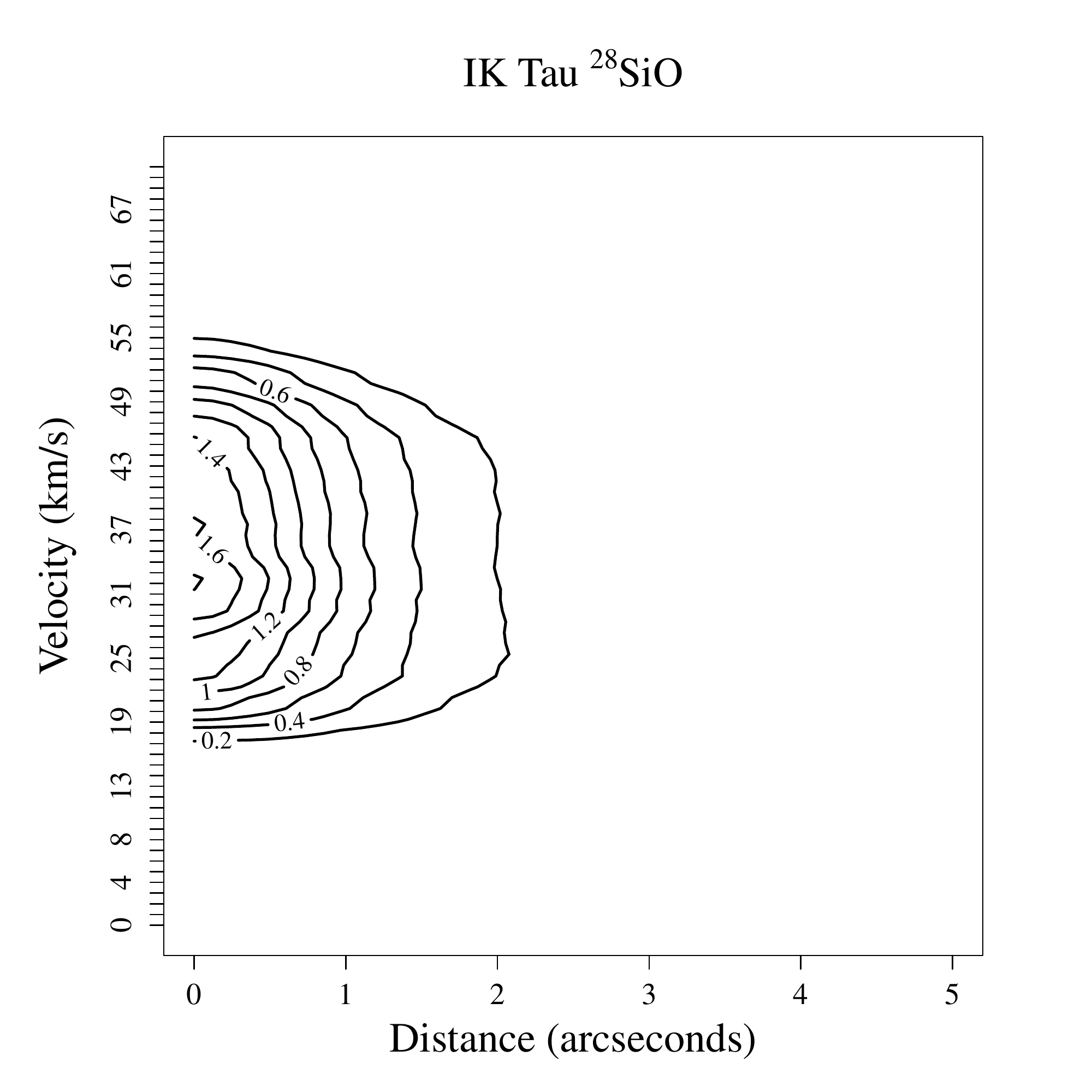}
\includegraphics[width=.48\textwidth]{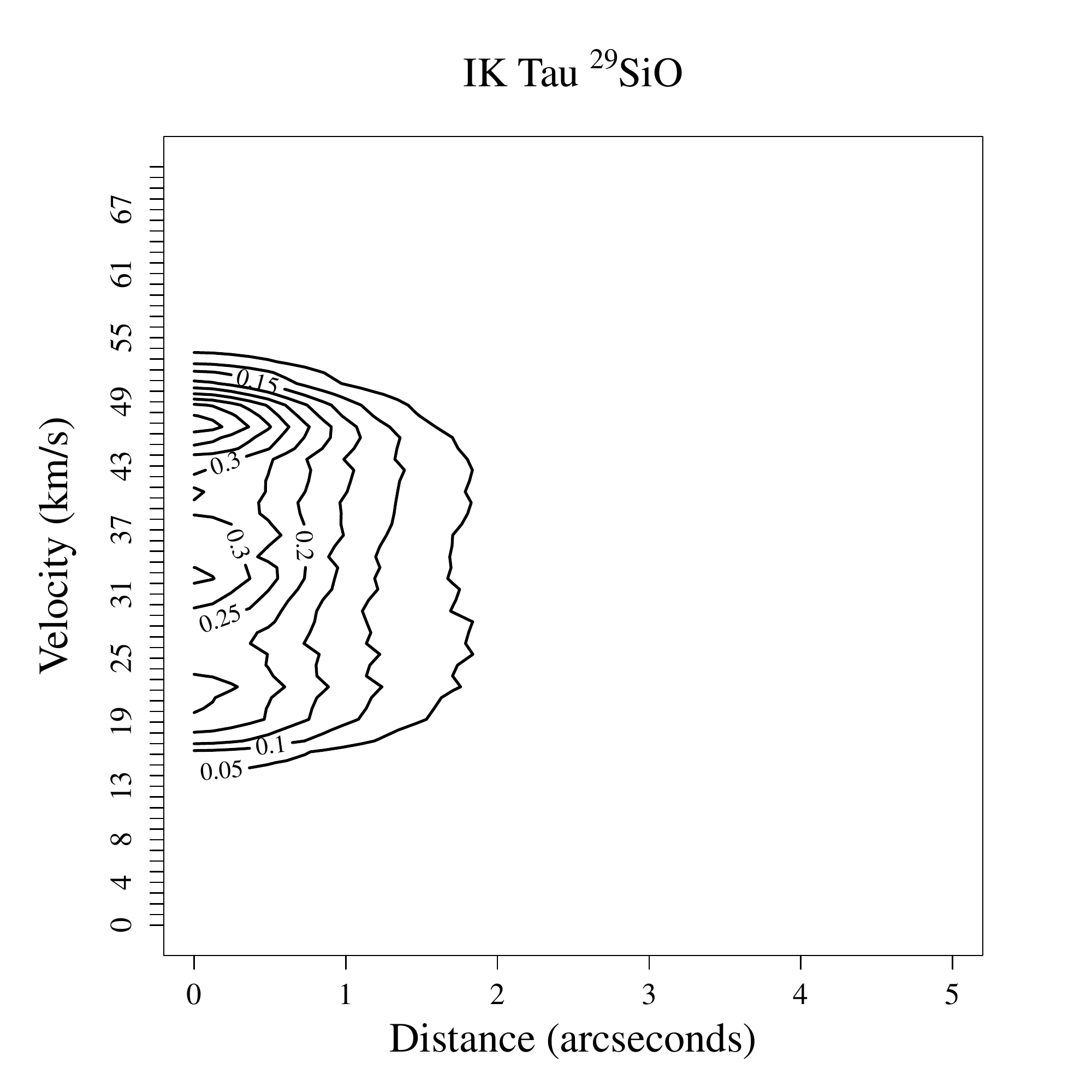}
\includegraphics[width=.48\textwidth]{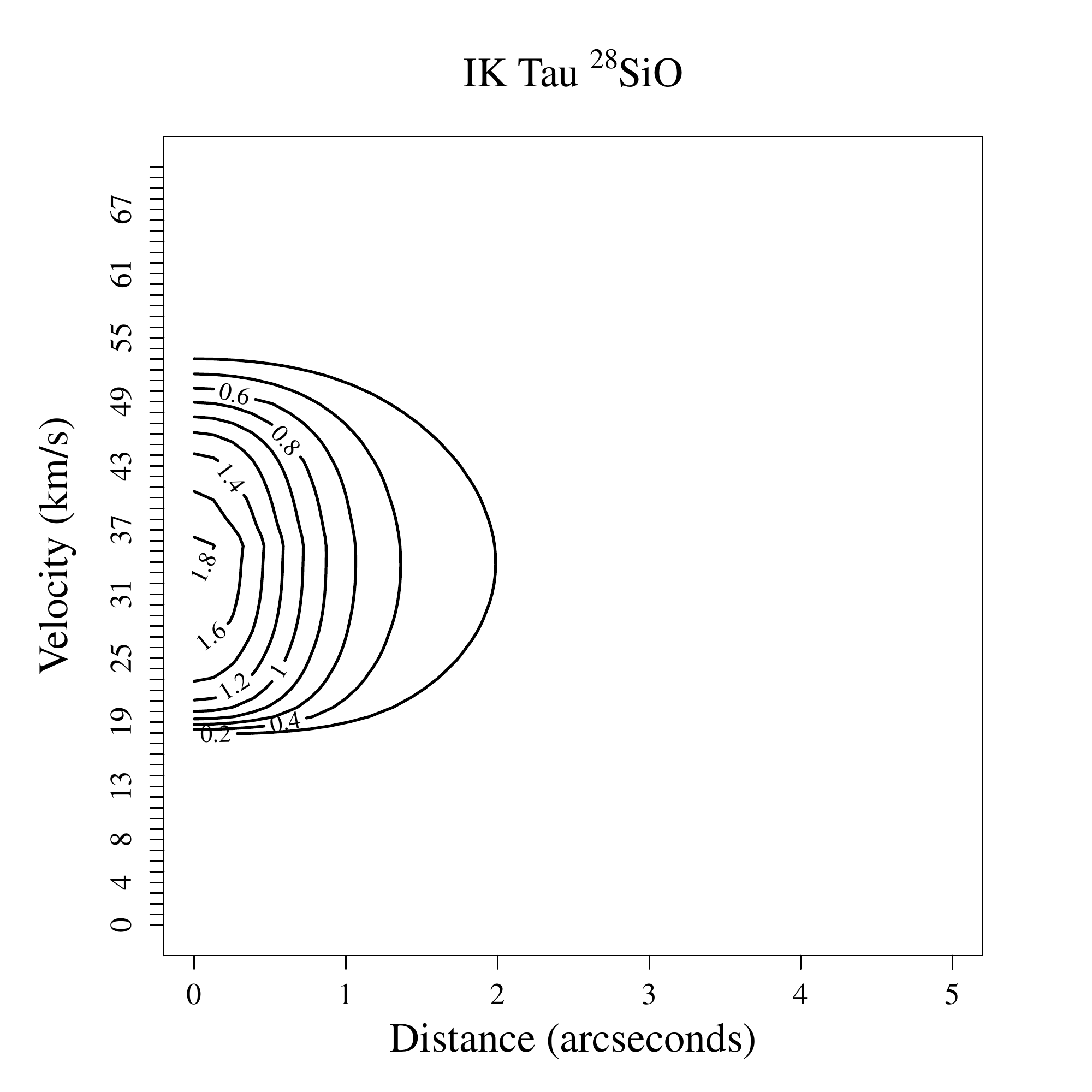}
\includegraphics[width=.48\textwidth]{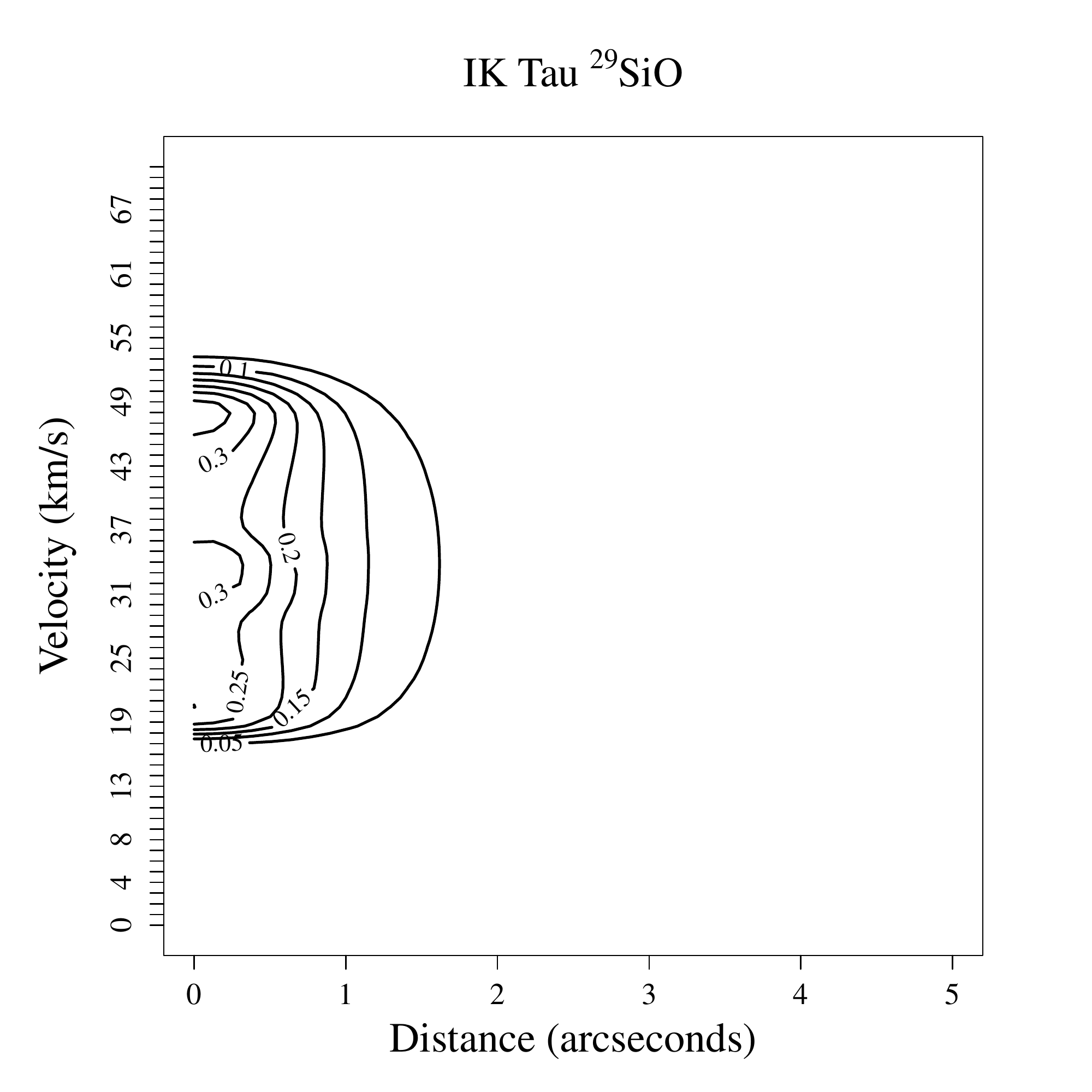}
\end{center}
\caption
{
Azimuthally averaged emission as a function of distance and LSR velocity for IK Tau. 
Observations are represented on the top and results of the radiative transfer model on the bottom.  
Both isotopologues are shown: $^{28}$SiO on the left and $^{29}$SiO on the right.
Countour labels are denoted in units of Jy~beam$^{-1}$.
}
\label{fig:iktaumods}
\end{figure*}

\begin{figure*}
\begin{center}
\includegraphics[width=.48\textwidth]{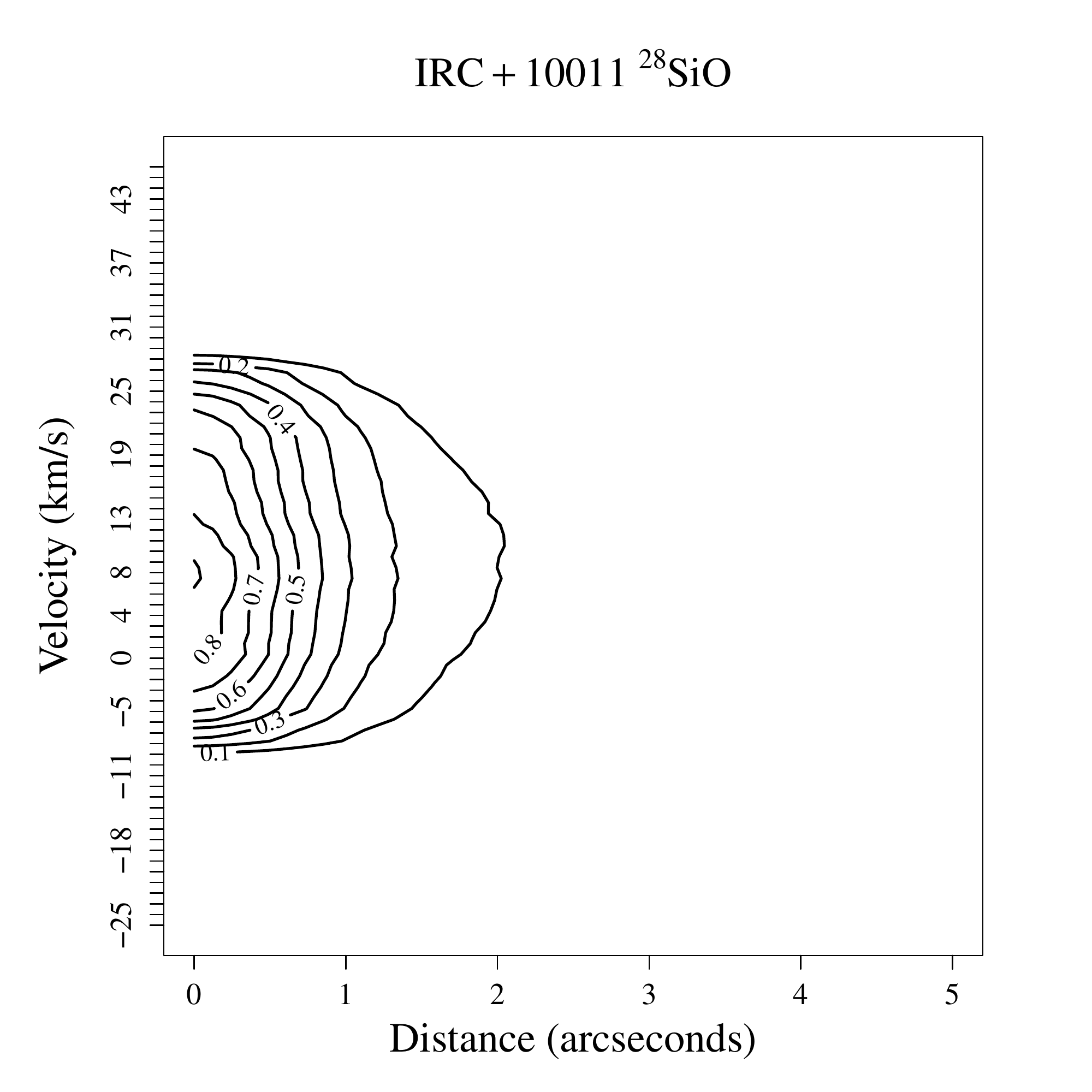}
\includegraphics[width=.48\textwidth]{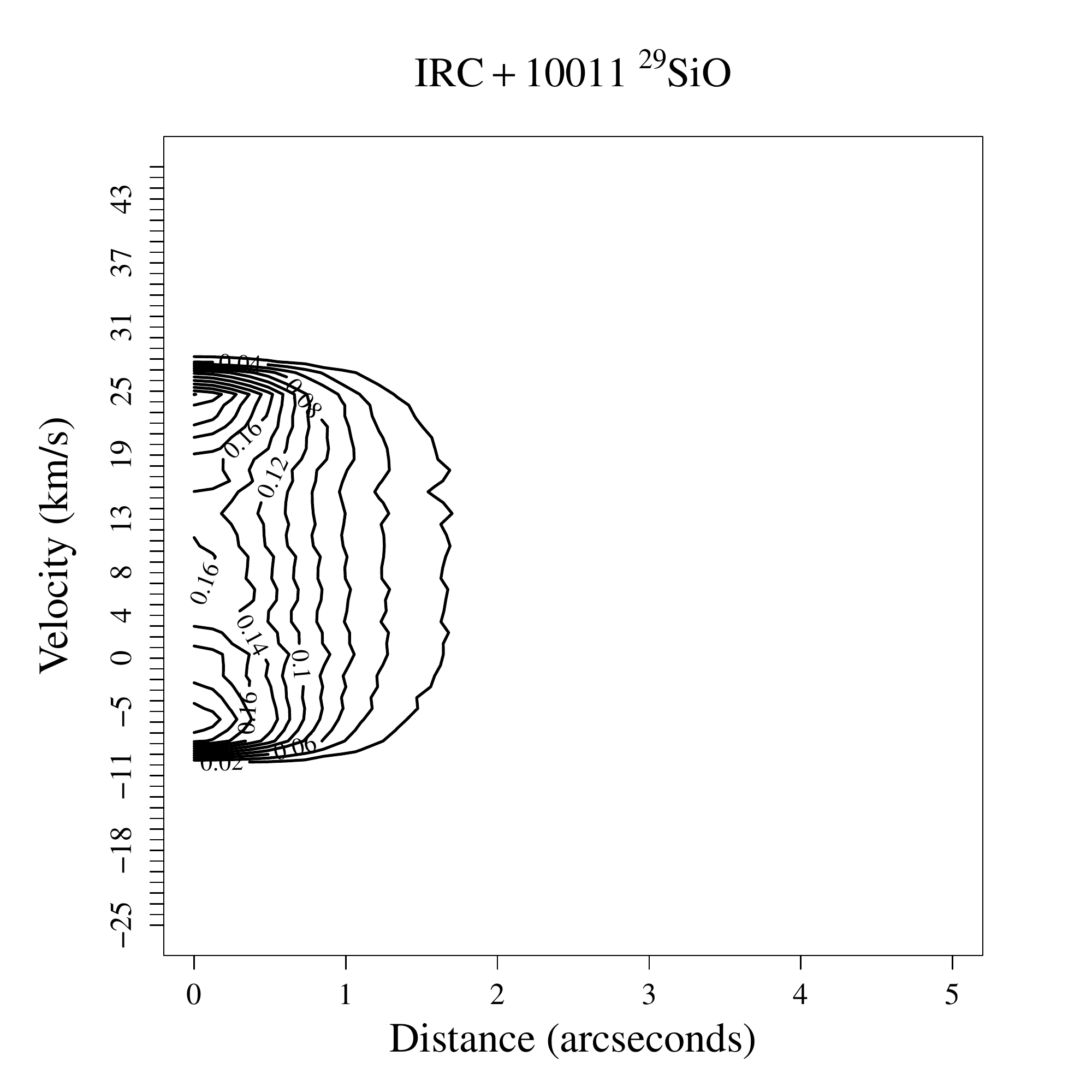}
\includegraphics[width=.48\textwidth]{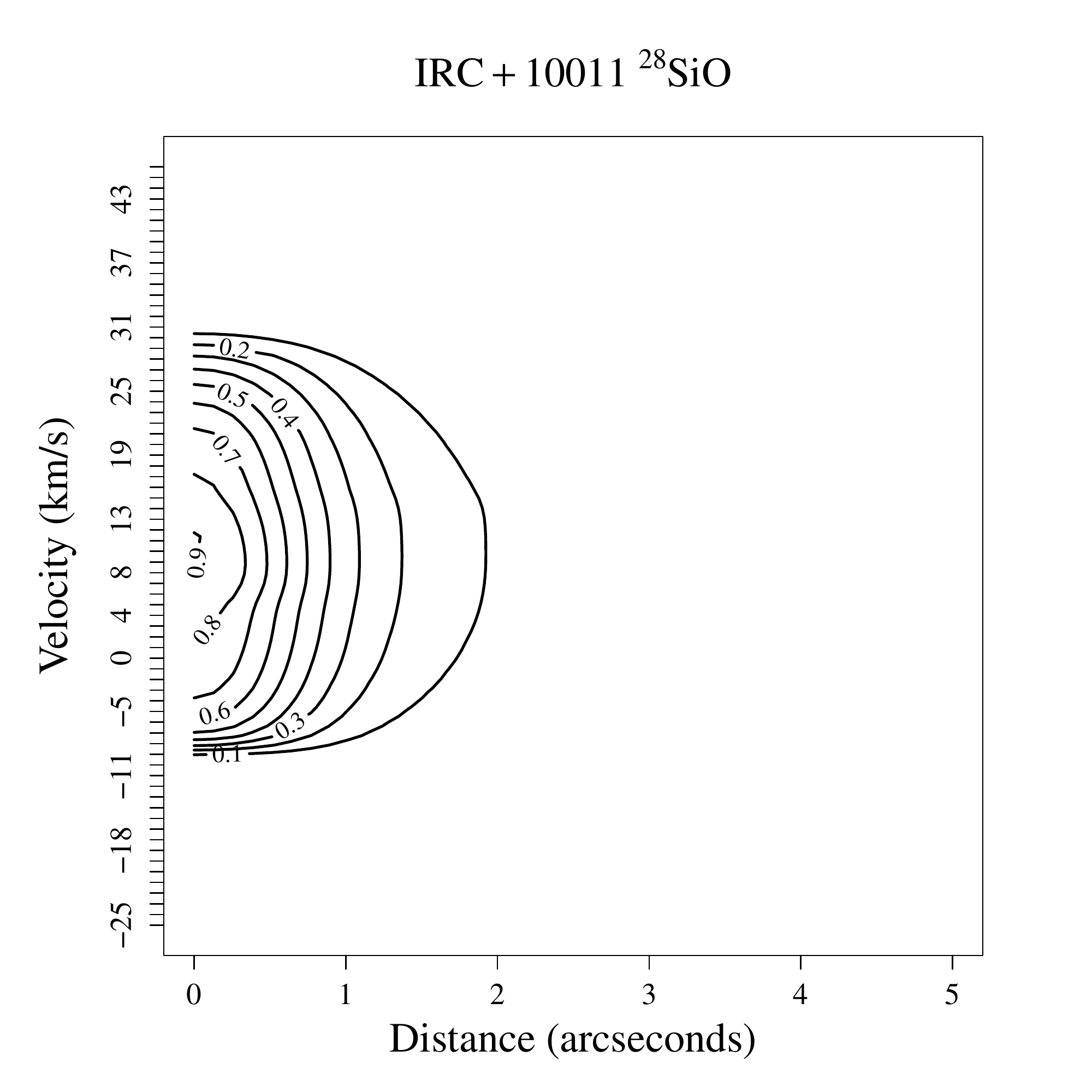}
\includegraphics[width=.48\textwidth]{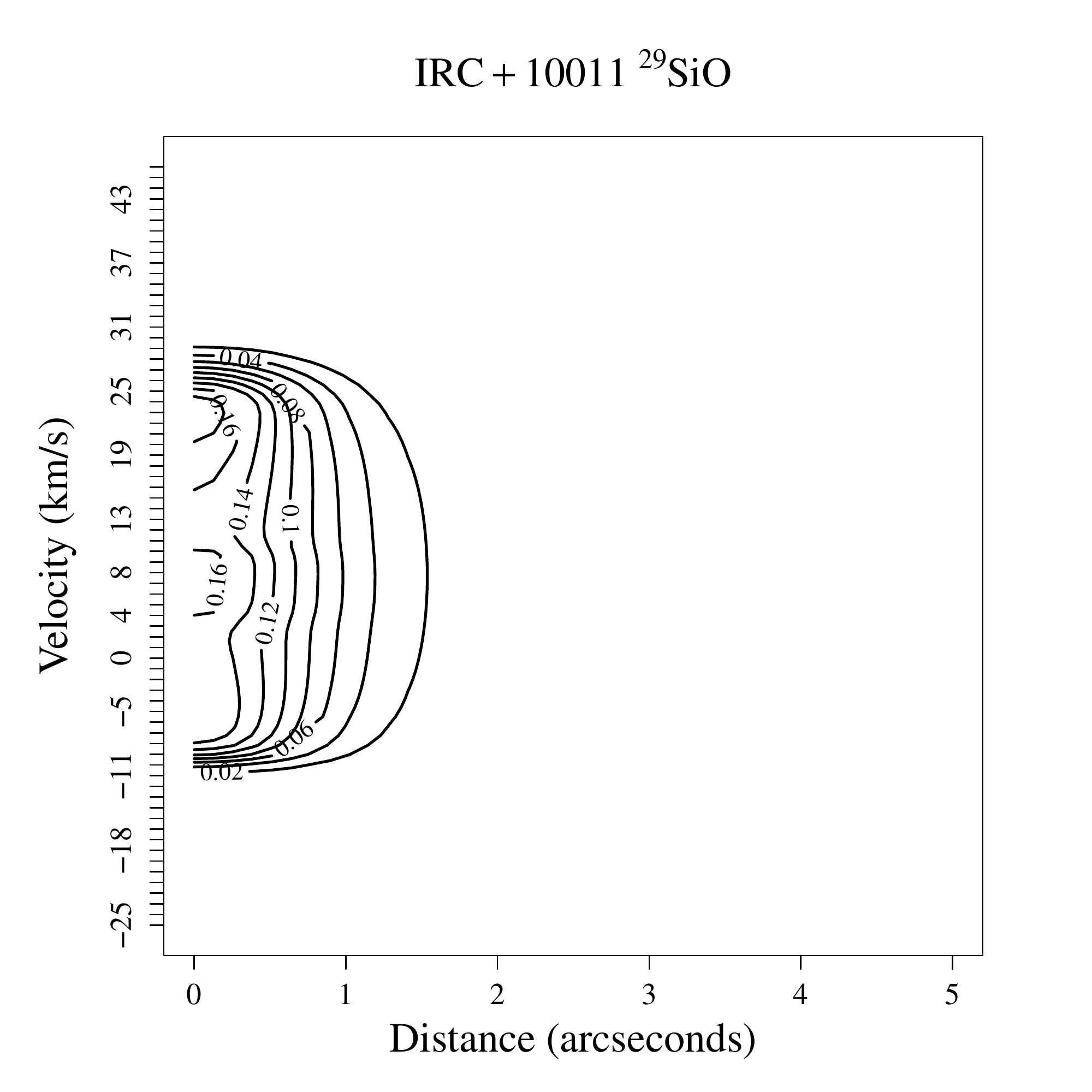}
\end{center}
\caption
{
Azimuthally averaged emission as a function of distance and LSR velocity for IRC+10011. 
Observations are represented on the top and results of the radiative transfer model on the bottom.  
Both isotopologues are shown: $^{28}$SiO on the left and $^{29}$SiO on the right.
Countour labels are denoted in units of Jy~beam$^{-1}$.
}
  \label{fig:ircmods}
\end{figure*}

From the observations we calculated azimuthal averages of the emission
and then computed an extensive number of models for the physical parameters of the 
circumstellar envelope
to reproduce the azimuthally averaged emission maps.
The computation of the azimuthal averages allows us to gain S/N in the
outer parts of the map and to compare lines at all distances with
our molecular excitation and radiative transfer code.
The code was tailored to efficiently treat
optically thin and optically thick thermal SiO emission. 
The observations, together with the best-fitting model results, are shown in Figs. \ref{fig:iktaumods}
and \ref{fig:ircmods}. 

The criteria used to determine the goodnes of fit to a 
model are 
\begin{itemize}
\item To fit both isotopic species with the same parameters varying only their relative abundance, assuming that
photodissociation is not isotope selective.
\item To obtain a reasonable and similar value for the $^{28}$SiO/$^{29}$SiO ratio for both objects.
\item To take previously published kinetic temperature and expansion velocity laws as starting points.
\item To fix mass-loss rates, terminal velocities, and distances to agree with previously 
published values.  
\end{itemize}

We considered that the density is a function of the radius, mass-loss rate, and expansion velocity following the continuity equation

\begin{equation}
\centering
\rho =  \frac{\dot{M}}{4 \pi r^2 V_e}
\label{dens}
\end{equation}

\noindent where $\dot{M}$ is the mass-loss rate, $V_e$ is the expansion velocity, and $r$ is the distance from the
central star.

The kinetic temperature profiles were determined using the following equation:

\begin{equation}
T_k = T_0 \left( \frac{2 \times 10^{15}}{r} \right)^{\alpha} + 5
\label{temp}
,\end{equation}

\noindent where $T_0$ and $\alpha$ are free parameters that determine the magnitude and slope of the
kinetic temperature.   

The velocity profile has been parameterised using a simplified law \citep{Lamers1999}. 
We used the following expression:

\begin{eqnarray}
\label{vexp}
V_e = V_{0} \hspace{1.8cm} \hspace{2cm} 
\textrm{if}~r~<~R_{d}  ,\\   
V_e = V_{0} + (V_{\infty} - V_{0}) \left( 1 - \frac{ R_{d}}{r} \right)^{\beta}   \hspace{0.5cm} \textrm{if}~r~\geq~R_{d}
\label{vexp2}
,\end{eqnarray}

\noindent where $V_0$ is the velocity before the gas starts to accelerate, which is set to 1 km~s$^{-1}$, $V_{\infty}$
is the terminal expansion velocity, 
and $R_{d}$ is the distance from the star at which dust grains form, which we treated as a free parameter that
affects the innermost portion of the envelope, where little is known about the dynamics.

To describe the SiO relative abundance profile
we used a simplified parameterisation

\begin{equation}  
X = (X_0 - X_{x}) \frac{R_x^{2}}{r^2+R_x^2}   
,\end{equation}

\noindent where $X_{x}$ is the final abundance left after SiO depletion in the inner envelope, and 
$X_O$ is treated as a free parameter that represents a reference abundance such that 
the initial value for $X$ is $X_0$ - $X_{x}$.
We assumed that 85\% of the gas-phase SiO is condensed into dust grains for both objects,
following \cite{Bujarrabal1989}.
The SiO abundance distribution is nonetheless expected to be very complicated because of
its strong dependence on dust parameters, in particular the dust mass loss rate \citep{Gonzalez2003}.
As can be seen, 
$R_x$ is the distance from the central star at which the abundance falls to $\frac{1}{2}$($X_0$ - $X_{x}$).

\begin{figure*}
\begin{center}
\includegraphics[width=.32\textwidth]{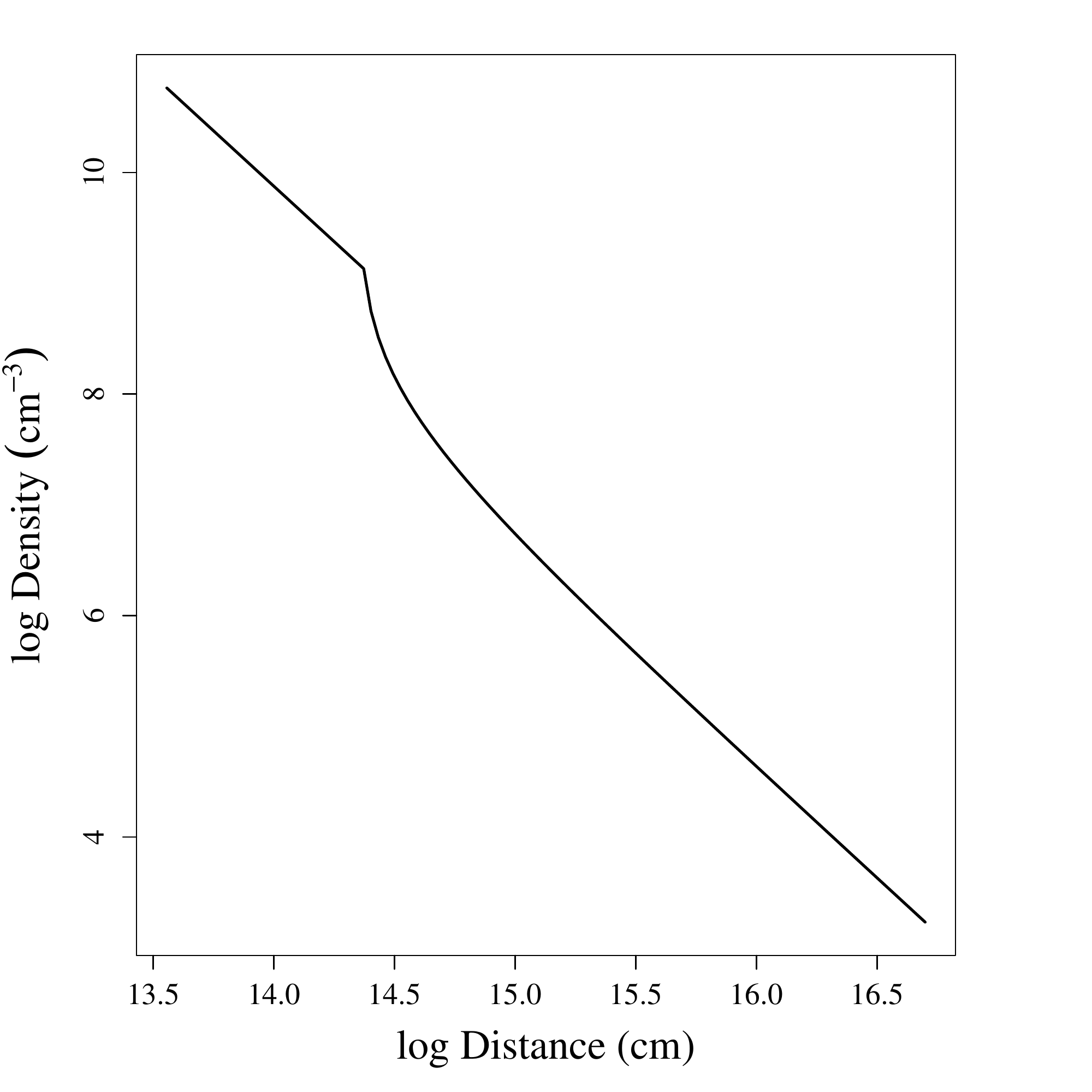}
\includegraphics[width=.32\textwidth]{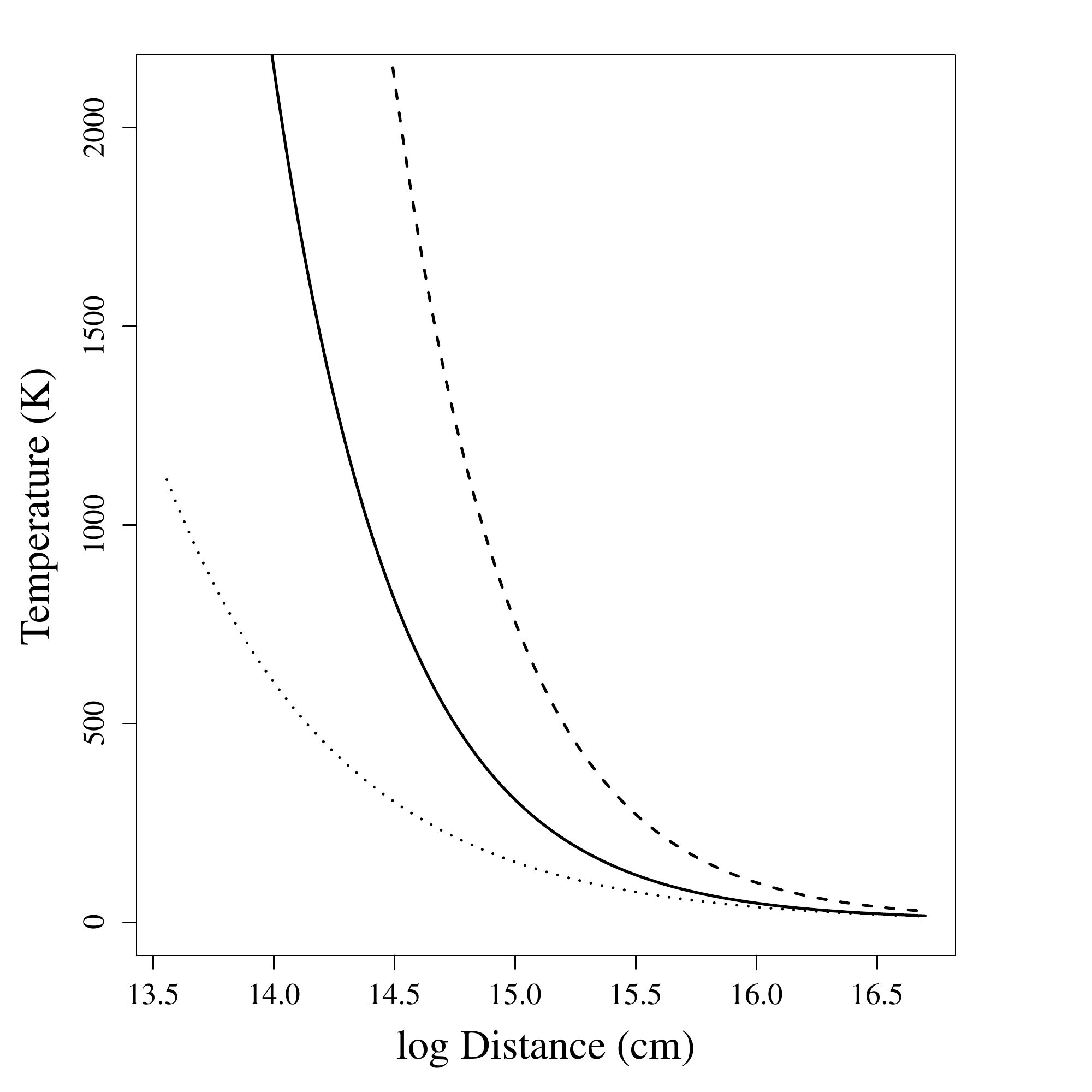}
\includegraphics[width=.32\textwidth]{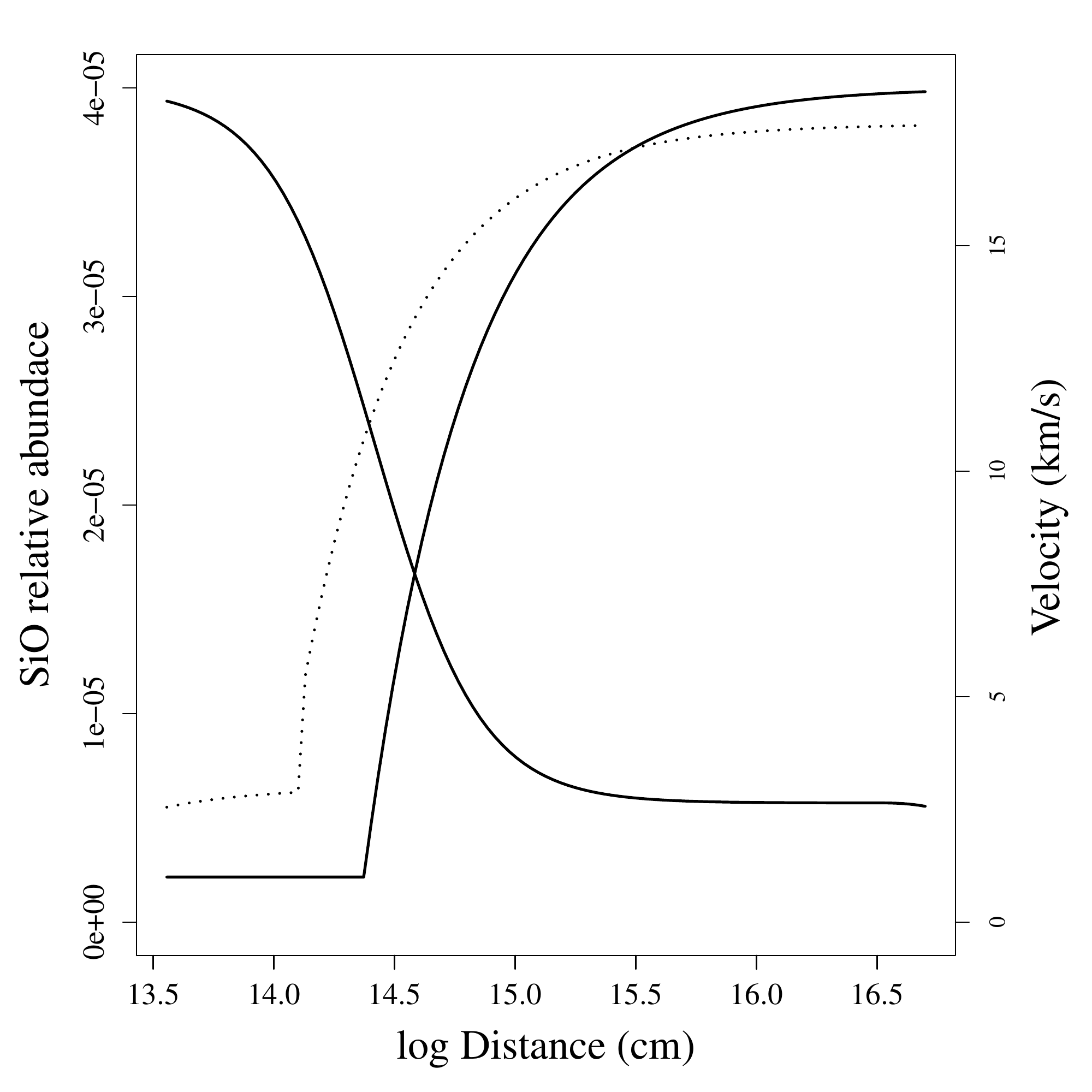}
\end{center}
\caption
{
Total density, kinetic temperature, molecular abundance, and velocity profile used for the models of Fig. \ref{fig:iktaumods}
for IK Tau.
We also show for comparison the temperatures obtained by \cite{Bujarrabal1989} (dashed line) and \cite{Decin2010}
(dotted line). In the velocity plot we include, for comparison, the velocity by \cite{Decin2010} (dotted line).
The velocity stays constant at 1 km~s$^{-1}$ up to the distance $R_d$, and then
gradually increases as described in Eqs. \ref{vexp} and \ref{vexp2}. 
}
  \label{fig:iktauparams}
\end{figure*}

\begin{figure*}
\begin{center}
\includegraphics[width=.32\textwidth]{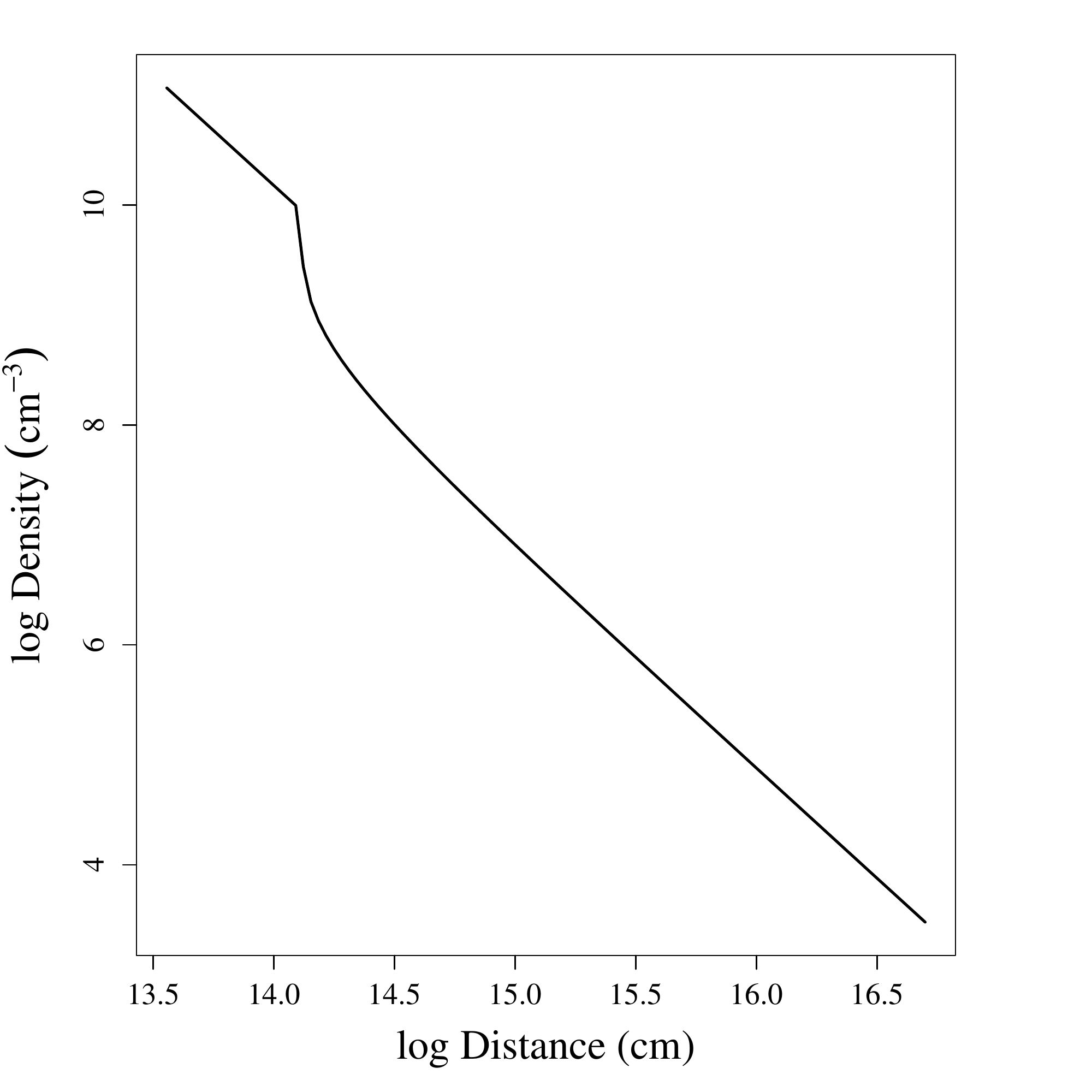}
\includegraphics[width=.32\textwidth]{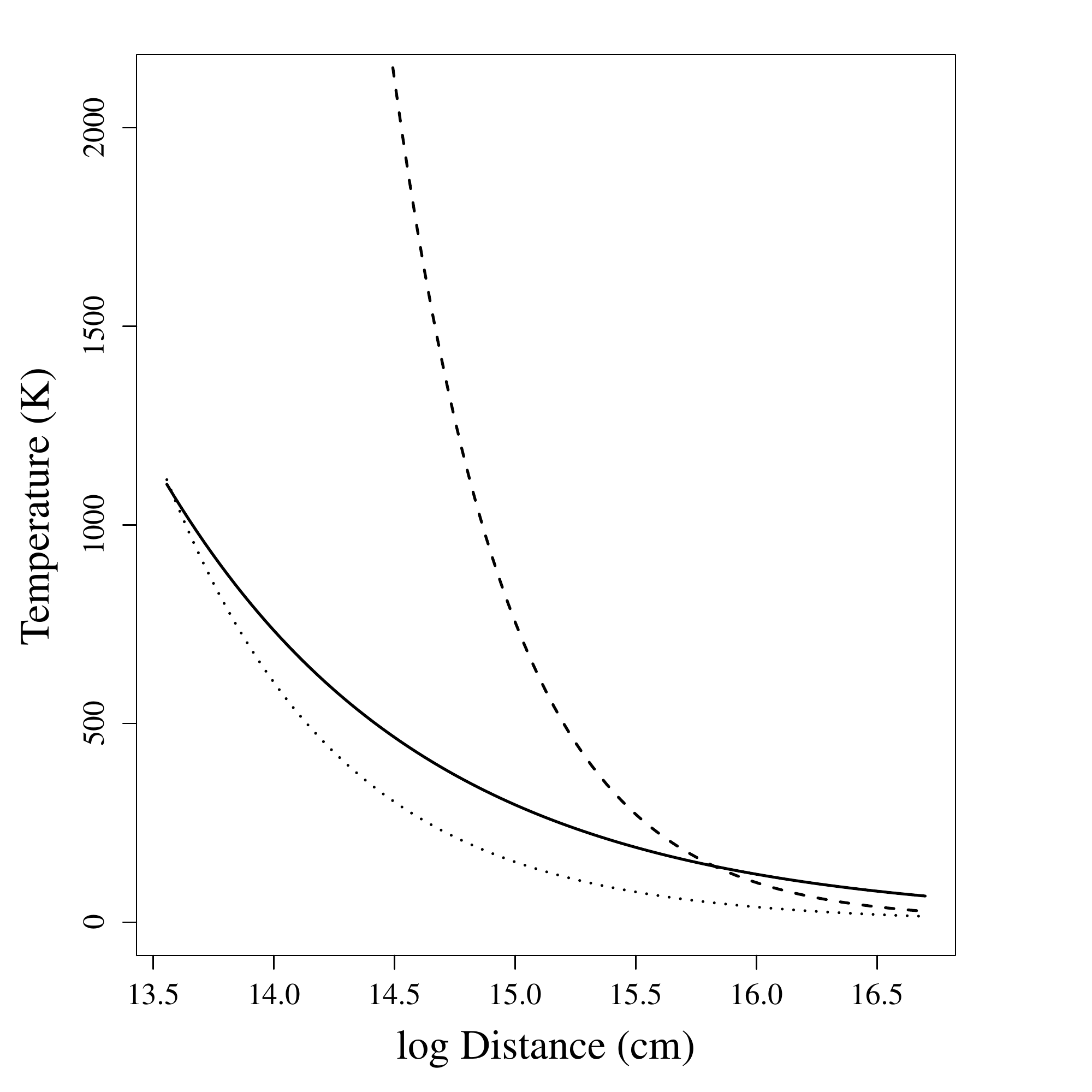}
\includegraphics[width=.32\textwidth]{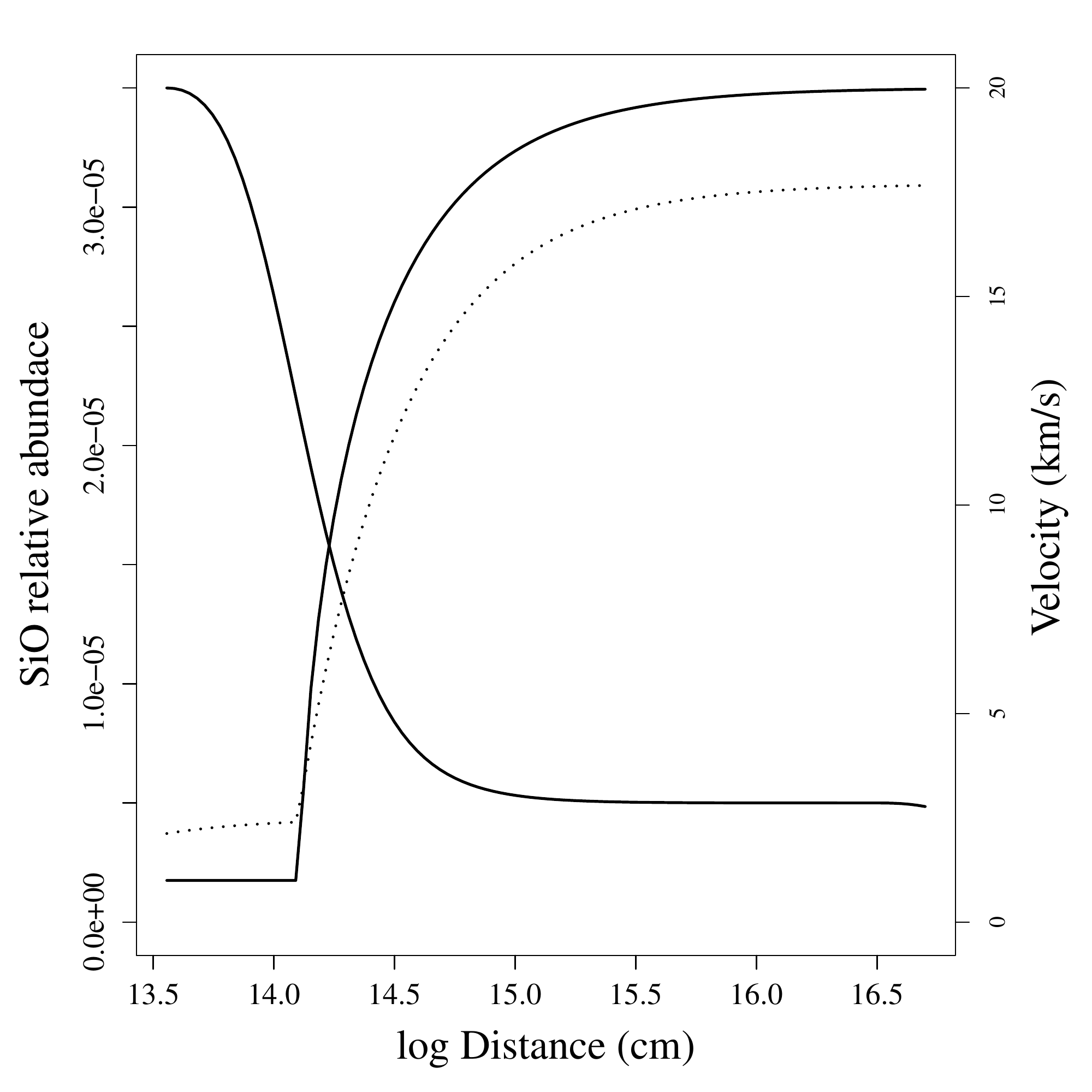}
\end{center}
\caption
{
Total density, kinetic temperature, molecular abundance, and velocity profile used for the models of Fig. \ref{fig:ircmods}
for IRC+10011.
We also show for comparison the temperatures obtained by \cite{Bujarrabal1989} (dashed line) and \cite{Decin2010}
(dotted line) for IK Tau. In the velocity plot we include, for comparison, the velocity by \cite{Decin2010}
for IK Tau (dotted line).  
The velocity stays constant at 1 km~s$^{-1}$ up to the distance $R_d$, and then gradually
increases as described in Eqs. \ref{vexp} and \ref{vexp2}. 
}
  \label{fig:ircparams}
\end{figure*}

The values of the fixed parameters adopted for each object are given in Table \ref{tab:fixedparams}.
The value for the distance to IK Tau is taken
from the Gaia Data Release 2 \citep{Gaia2018}, while that of IRC+10011 is taken from \cite{Vinkovic2004} and
references therein. We adopt the mass-loss rates reported in \cite{Lucas1992} for both objects, which
have been corroborated by other more recent works (see e.g. \citealp{Decin2018}). Stellar temperatures used to calculate
stellar excitation are taken from \cite{Decin2018} for IK Tau and \cite{Vinkovic2004} for IRC+10011.
Terminal expansion velocities were taken from \cite{Velilla2017} and \cite{Vinkovic2004} for IK Tau and IRC+10011,
respectively.
We fixed ${^{28}}$SiO/${^{29}}$SiO = 20 for
both objects, similar to the galactic
value and in agreement with the values obtained by \cite{Monson2017}.

The free parameters and their values are listed in Table \ref{tab:freeparams}. 
The density, kinetic temperature, abundance, and expansion velocity for the models of both
sources are shown in Figs. \ref{fig:iktauparams} and \ref{fig:ircparams}.
For both objects we found kinetic temperature laws somewhat in between those of \cite{Bujarrabal1989}
and \cite{Decin2010}. Our results are plotted in Figs. \ref{fig:iktauparams} and \ref{fig:ircparams}.
We note that these laws were obtained for IK Tau, but we also included them in the 
plots for IRC+10011 for comparison.
The kinetic temperature found by \cite{Bujarrabal1989} is higher in general, but decreases steeply as $r^{-0.9}$,
while that of \cite{Decin2010} decreases more slowly as $r^{-0.6}$.


While having observations of two isotopologues for two relatively similar sources 
helps us to more strictly constrain the parameters in our models, it also results in complications in the achievement of
satisfactory model fits and in the calculation of uncertainties.
The criteria to obtain the parameters of the model that best fit the observations were the following:
From the azimuthally averaged emission maps
that resulted from the observations we calculated the r.m.s. value of the noise. It is calculated outside the emission
region in emitting LSR velocities to account for cleaning errors and we call it $\sigma$.

We computed a set of models representing different scenarios,  such as slow/fast acceleration, early/late SiO
depletion,  and we obtained maps of azimuthally averaged emission.
We visually inspected the model maps and compared these maps with the azimuthally averaged emission maps 
that resulted from the observations 
to infer which models reproduced the observations better, and to
identify problems and artefacts that would be hard to identify numerically.
Additionally, we subtracted the observed map
from each resulting model map, thereby obtaining a residual map.
In this residual map we
calculated the r.m.s. in the zones with emission, and also inspected the maps visually to identify zones
of large residual value and/or other systematic errors. 
The model we chose as the most adequate is the one that results in the residual map with the 
minimum r.m.s. and minimum residual value. 
This process is done simultaneously for both $^{28}$SiO and $^{29}$SiO emission because the set
of parameters has to reproduce the emission for both isotopic species varying only their relative
abundance to H$_2$. Some compromise had to be made because while the variation of
a given parameter might improve the fit for an isotopic species, at the same time it could worsen
the fit for the other. 
 
Once we had the general idea of the scenario, we then computed another set of models with slighter variations
of the parameters and chose the most adequate model in the same manner that we did  
in the previous step and also for both isotopologues simultaneously. We established as the most
adequate models those that satisfy the following conditions for both isotopic species: the r.m.s. in
the emission region stays below 3$\sigma$, the maximum residual has an absolute value below 5$\sigma$, and
the size of the region with this maximum residual stays smaller than the beam size. 

Once we had established the set of parameters that best reproduced the observations, 
to estimate the uncertainties in these parameters (see Table \ref{tab:freeparams}) we varied one
parameter at a time while keeping the others constant and obtained residual maps. 
Once the map contained residuals with values exceeding 7 $\sigma$ or had a r.m.s. in the
region with emission exceeding 3$\sigma$ for either of the isotopic species, we regarded that model as
no longer acceptable and reported the uncertainties as the maximum variation the parameter can take; we also
always  visually inspected the maps. 
A more detailed description of the method is given in Appendix \ref{err}.

Another source of uncertainty is the fact that we constrained the kinetic temperature law by a single SiO line,
however our results for IK Tau agree with those of \cite{Bujarrabal1989} (see Fig. \ref{fig:iktauparams}). These authors used
CO ($J=1-0$) and SiO ($J=2-1$) lines to constrain the kinetic temperature profile. Unfortunately, 
to our knowledge, similar information 
for IRC+10011 is not available.


\begin{table*}
    \centering
    \caption
    {
    Fixed parameters used for the models.
    }
    \label{tab:fixedparams}
    \begin{tabular}{@{}lcccccc@{}} 
        \hline
        Object & $\dot{M}$ (M$_{\odot}$ yr$^{-1}$) & Distance (pc) &  
        $\frac{^{28}\textrm{SiO}}{^{29}\textrm{SiO}}$  & $V_{\infty}$ (km~s$^{-1}$) \\

        \hline      \\[0.1cm]

        IK Tau  & 5 $\times$ 10$^{-6}$  & 285  & 20  & 18.5    \\
        IRC+10011 & 1 $\times$ 10$^{-5}$  & 650  & 20  & 20      \\
        \hline
    \end{tabular}
\end{table*}

\begin{table*}
    \centering
    \caption
    {
    Free parameters used for the models.
    }
    \label{tab:freeparams}
    \begin{tabular}{@{}lcccccc@{}} 
        \hline
         Object  & $R_d$ ($10^{14}$~cm)   & $\beta$  & $T_0$ (K) & $\alpha$ & $R_{X}$ ($10^{14}$~cm)  & $X_0$(SiO) ($10^{-5}$) \\

        \hline      \\[0.1cm]

IK Tau & 2.2$^{+0.1}_{-0.5}$ & 1.0$^{+0.1}_{-0.5}$ & 168$^{+19}_{-8}$& 0.85$^{+0.70}_{-0.20}$ & 2.6$^{+0.7}_{-0.4}$ & 4.0$^{+0.3}_{-0.1}$ \\

      \\[0.05cm]

IRC+10011& 1.2$^{+0.1}_{-0.5}$  & 0.6$^{+0.3}_{-0.1}$ & 220$^{+30}_{-45}$ & 0.40$^{+0.25}_{-0.10}$ & 1.0$^{+0.3}_{-0.3}$ & 3.5$^{+0.2}_{-0.1}$ \\
        \hline
    \end{tabular}
\end{table*}

\begin{figure*}
\begin{center}
\includegraphics[width=.45\textwidth]{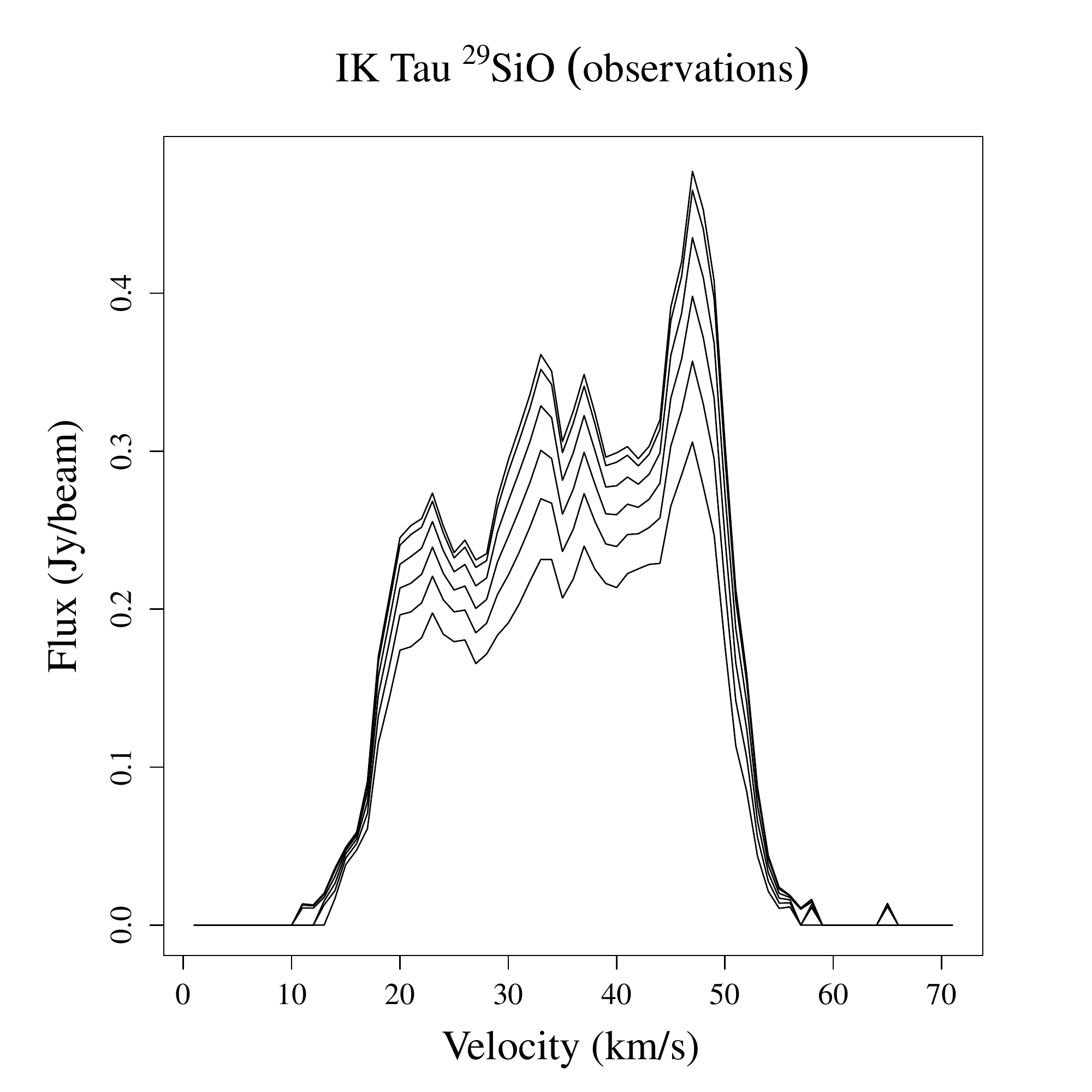}
\includegraphics[width=.45\textwidth]{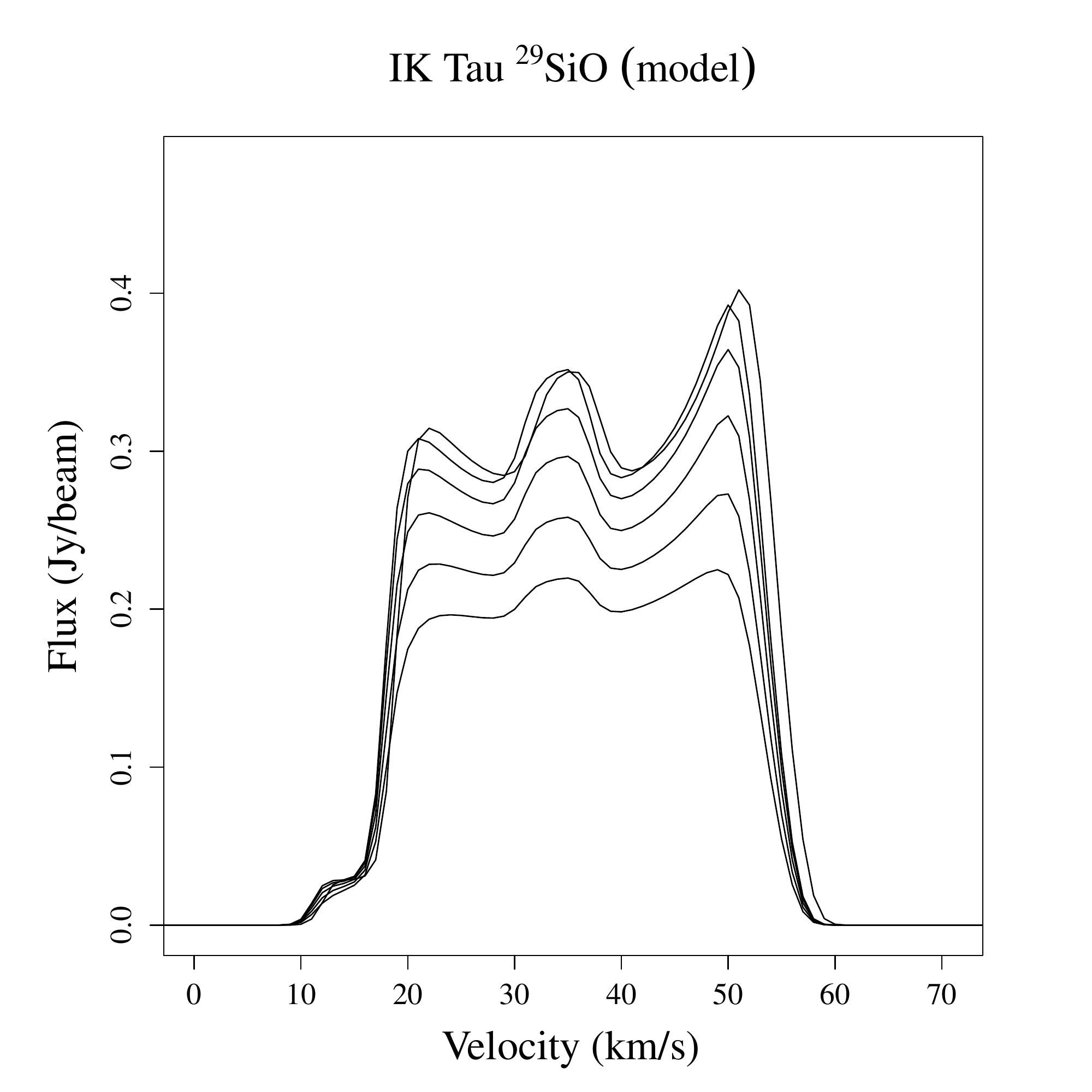}
\end{center}
\caption
{ 
Spectra of the innermost regions of the envelope. The azimuthally averaged observed emission is shown to the left 
and the model to the right. Six positions are shown separated by 0''.125; the innermost is the most intense.
The model reproduces to some extent the three peaks and the self-absorption that happens 
in the blue-shifted side of the spectra.
}
  \label{fig:spectra}
\end{figure*}

As we can see in Figs. \ref{fig:iktauparams} and \ref{fig:ircparams}, the best fits are those which
result from a gas that is still accelerating in the inner region of the wind and that has not acquired
its terminal expansion velocity as quickly as it was previously believed
(see e.g. \citealp{Olofsson1982}, \citealp{Morris1983}, \citealp{Schonberg1988}). 
We find that there is a good correlation
between the decrease in the abundance of the SiO molecule and the gas acceleration. 
If the SiO is removed from the gas phase as a result of grain growth, 
radiation pressure from the star then
acts more efficiently on larger dust grains. Owing to a strong coupling between gas and dust, gas then
accelerates along with the dust. 
Therefore, a correlation between gas-phase SiO depletion and gas acceleration might
be a sign of SiO condensation into dust grains.
Our results indicate that it is reasonable to assume that $\sim$ 85\% percent 
of the SiO molecular gas condenses into dust grains, and
the remaining SiO photodissociates very far from the star.
We also note that gas acceleration
and SiO depletion happen slightly faster for IRC+10011 than for IK Tau. 
This is understandable if we take into
account the larger mass-loss rate, and hence larger density, of the former: under this condition 
grain growth happens more quickly, and therefore the dust accelerates faster and SiO
depletion also happens faster.
For this object we also assumed that the value of the temperature is in general lower, which also
favours the condensation of dust grains.

After the molecular gas is condensed 
into dust, at much larger
distances from the star, photodissociation removes the remaining SiO in the gas phase. 
We assumed that photodissociation sets in at a distance beyond 3~$\times$~10$^{16}$~cm for both objects
following the results of \cite{Li2016}. 
Using a large gas-phase chemical model of an AGB envelope
including the effects of CO and N$_2$ self-shielding,
along with an improved list of parent species derived from detailed modelling and observations,
concluded that the SiO
photodissociation region lies past
3~$\times~10^{16}$ cm for IK Tau.


In Fig. \ref{fig:spectra} 
we show azimuthally averaged
spectra obtained from observations of $^{29}$SiO in IK Tau and spectra that results from the modelling.
We can see two interesting features of the obervations
that were reproduced in the models to some extent. 
One is is self-absorption and the other is a triple-peaked structure of
the azimuthally averaged central spectra. 
Self-absorption
happens when a region of
relatively low excitation temperature lies somewhere along the line
of sight to a hotter region with the same radial velocity. 
The other feature, the triple-peaked central spectra, means that
we find local maxima at 
central velocities, as
well as in the blue and red regions of the spectra. 
The blue and red peaks are very common in optically thin emission of outflows because
we see the approaching and receding wind. The 
central peak
appears
if we have a high SiO abundance in the inner zone with slow-moving gas 
causing an emission excess. 
The three peaks can clearly be seen as well as the self-absorption in the blue-shifted
side of the spectra. The figures include the emission at six offsets from the central star separated by 0''.125
to make the effects more visible.

\section{Discussion} \label{sec:discu}

We have been able to reproduce, quantitatively and qualitatively, the emission
from the inner envelope of IK Tau and IRC+10011 within the expected uncertainties,
including the effects of self-absorption, as can be seen especially in the blue part of
the spectrum of the $^{29}$SiO emission. 
We also 
reproduce in the models the triple-peaked structure of the azimuthally averaged central spectra.
The central peak
appears
because we have a high SiO abundance in the inner zone with slow-moving gas 
that causes an emission excess.
This supports our hypothesis that
the inner part of the envelope is richer in gas-phase SiO and is moving at lower velocities before the
gas accelerates and strong SiO depletion occurs. 

The sizes of the SiO emission we computed (Section \ref{sec:res}) differ to
those found by \cite{Gonzalez2003}. For IK~Tau these authors found a radius of the
SiO emission region of 2.5~$\times$~10$^{16}$~cm,
which is larger than the value we found by a factor of 4.6. 
It is worth noting that in their work the radius was set 
as a free parameter, and they reported the radius value that provided the best fit.   
We also note that they used the rather high mass-loss
rate of 3~$\times$~10$^{-5}$~$M_{\odot}~yr^{-1}$ compared to the value of 5~$\times$~10$^{-6}$~$M_{\odot}~yr^{-1}$ 
used by us.
Our values are, nonetheless, very similar to those found by \cite{Lucas1992}, taking into account
that they underestimated the distances to the sources.

The value of the terminal expansion velocity for IK Tau agrees with that found by 
\cite{Velilla2017}, who estimated an expansion velocity of 18.5 km~s$^{-1}$ from the line widths of the spectral
features arising from the outer envelope, where the gas is assumed to have
fully accelerated to its maximal velocity. The value for this terminal expansion velocity is 
higher than the previously found value of 17 km~s$^{-1}$ \citep{Lucas1992}.

\section{Summary and conclusions} \label{sec:conc}

Using high angular resolution images of SiO thermal emission in the circumstellar envelopes of
IK Tau and IRC+10011 we estimated the half-power diameters and fluxes as functions of velocity
assuming the emission region is well represented by a Gaussian disc. 
We conclude that the fact that the variation of the diameter of the emission region 
is small is a sign that gas
is still accelerating in the inner envelope where dust grains are forming. 
Assuming spherical symmetry we computed azimuthal averages.
We were able to estimate the physical parameters that reproduce the
emission using a molecular excitation and ray-tracing code tailored to treat SiO thermal
emission. We used as starting points for our models previously found 
values of parameters such as mass-loss rate, distance,
stellar temperature, and terminal velocity. We used temperature laws, expansion velocity
laws, and isotopic ratios, which are in agreement with values previously reported in the literature.   
We conclude there is a strong coupling between the depletion of gas-phase SiO and gas acceleration
in the inner region of the envelope. Both the SiO depletion
and gas acceleration could be explained by the condensation of SiO into dust
grains.
The difference in acceleration and SiO depletion
between the two objects can be explained in terms of their mass-loss rates and temperatures. SiO depletion
and gas acceleration occur faster in the higher mass loss, lower temperature object because 
the appropriate conditions for dust formation appear sooner and at shorter distances from the star.

\section*{Acknowledgements}
The research leading to these results has received funding from the European 
Research Council under the European Union's Seventh Framework Programme (FP/2007-2013) / ERC 
Grant Agreement n. 610256 NANOCOSMOS and by the Spanish MINECO, Grant FIS2012-32096.
This work has made use of data from the European Space Agency (ESA) mission
{\it Gaia} (\url{https://www.cosmos.esa.int/gaia}), processed by the {\it Gaia}
Data Processing and Analysis Consortium (DPAC,
\url{https://www.cosmos.esa.int/web/gaia/dpac/consortium}). Funding for the DPAC
has been provided by national institutions, in particular the institutions
participating in the {\it Gaia} Multilateral Agreement.
The authors thank the anonymous referee for comments and suggestions that improved the quality of the manuscript.
This research has made use of NASA's Astrophysics Data System.

\begin{appendix} 

\section{Obtaining the best-fit model and parameter uncertainties} \label{err}

\begin{figure}[ht]
\begin{center}
\includegraphics[width=.45\textwidth]{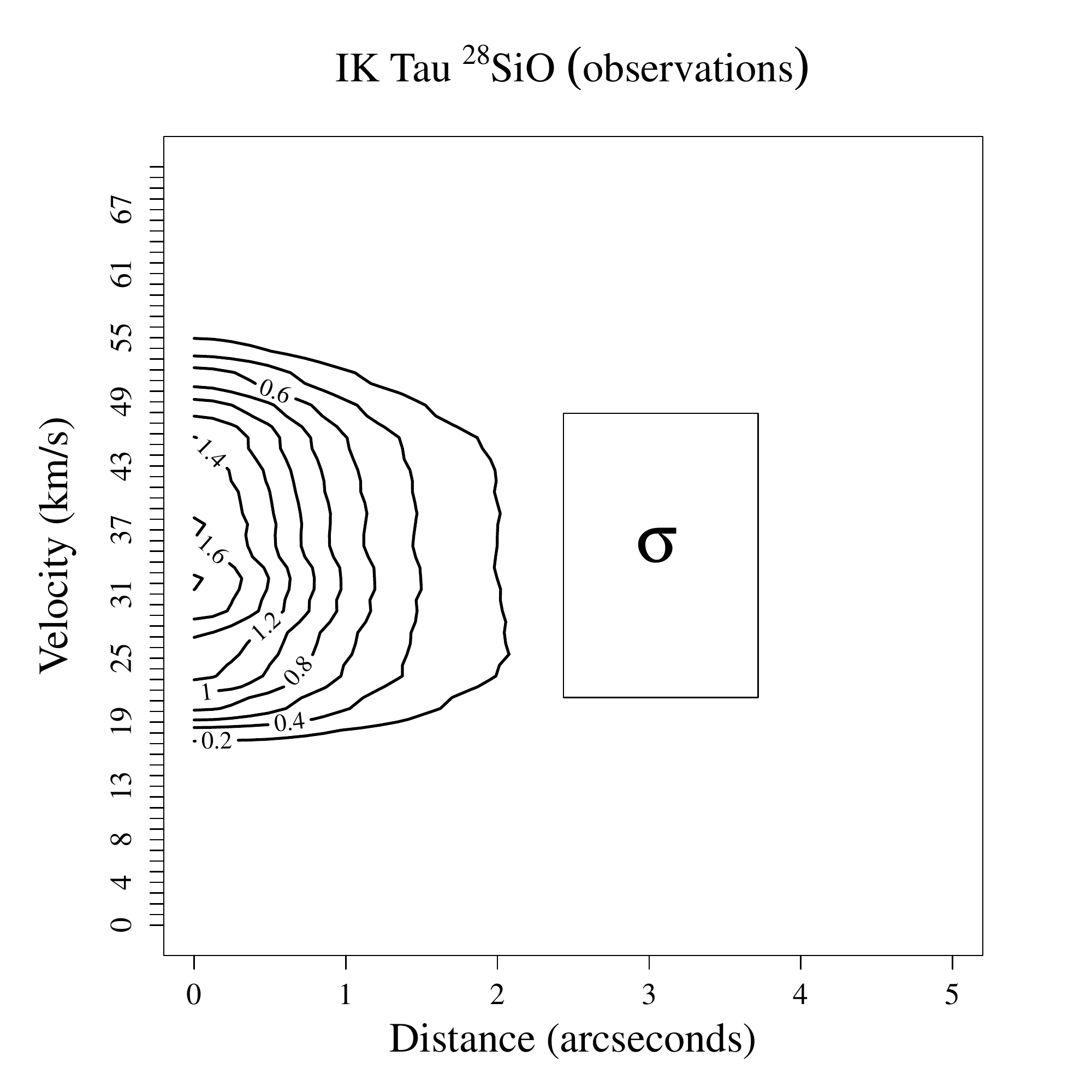}
\end{center}
\caption
{ 
Azimuthally averaged $^{28}$SiO emission as a function of distance and LSR velocity for IK Tau. 
The rectangle outside the contours 
represents the area where the r.m.s. ($\sigma$) was calculated. We chose this region to take into
account errors that could have resulted from the cleaning process.  
}
  \label{fig:rmsreg}
\end{figure}

We illustrate the method used to obtain the models that best fit the observations. In Fig. \ref{fig:rmsreg} we show the azimuthally averaged emission map obtained from the observations. The rectangle
outside the contours represents the area where the observational r.m.s. value of the noise is calculated. We chose
this region to take into account errors resulting from the cleaning process. We call this r.m.s. value, $\sigma$.  
For this particular case we have $\sigma$~$\sim$~0.07 Jy beam$^{-1}$.  

We compute a set of models varying the parameters for the abundance, expansion velocity, and kinetic
temperature laws to account for different scenarios, i.e. slow or fast acceleration, SiO depletion happening
sooner or later, etc.  
We compute azimuthally averaged maps of the emission from the models.
Besides visually inspecting and comparing each of the resulting models to the observations, we subtract from the model map 
the observed map and obtain a residual map. 
Our goal is to obtain a model that results in a residual
map with the minimum r.m.s. in the emission zone. We cannot rely solely on this criterion to discriminate 
between models, as it can average out
large residuals with the very small.
Therefore, we use as an additional
criterion that the zones with maximum residuals must have an absolute value below 5$\sigma$ and a size no more
extended than
the size of the beam.   

Once we find the general trends for the abundance, expansion velocity, and kinetic
temperature laws that best reproduce the observations, we start from there and vary each of 
the parameters slightly until we find the model that results in the residual map with the minimum r.m.s. and 
the minimum residual values.

\begin{figure}
\begin{center}
\includegraphics[width=.45\textwidth]{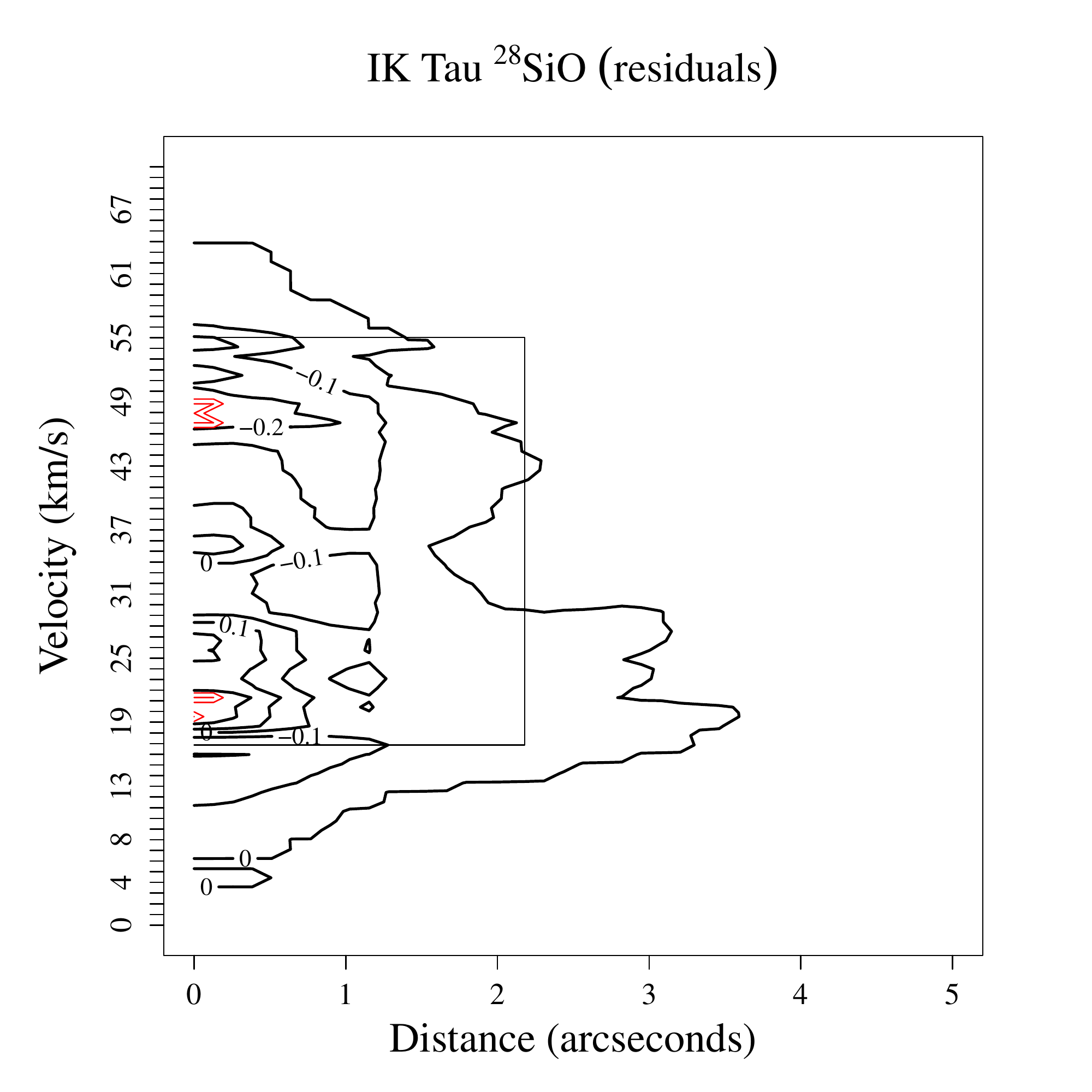}
\end{center}
\caption
{ 
Residual map that results from subtracting the observations from the model, for IK Tau, $^{28}$SiO emission. The 
r.m.s. of the residuals is calculated in the
area enclosed by the rectangle. 
The red contours illustrate the regions in which the maximum, in absolute value, residual is located.  
}
  \label{fig:diff}
\end{figure}

If we subtract the observations  
from the model, for the case of $^{28}$SiO emission shown in Fig. \ref{fig:iktaumods},  
we obtain the residual map shown in Fig. \ref{fig:diff}. When we
calculate the value of the r.m.s. of this map in the region with emission, represented by the rectangle, 
we obtain a value of  
$\sim$~0.09~Jy~beam$^{-1}$, which
is $\sim$ 1.3$\sigma$. 
This is done while also visually
inspecting the resulting models to avoid other systematic errors. For this particular case, the absolute value of the
largest residual is $\sim$ 0.28~Jy~beam$^{-1}$ (4$\sigma$) and is located in the regions enclosed
by the red contours.      

The fact that the same set of parameters has to satisfy the observations for both isotopologues, as we assumed that
photodissociation is not isotope selective, complicates the process, since variations of the parameters that improve
the fit for one of the isotopic species, might at the same time worsen it for the other. For this reason the modelling
process was carried out for both isotopologues simultaneously and some compromise had to be made.
The model we chose as the best-fit  results in a residual map with an r.m.s. below 3$\sigma$ and 
contains no
residual values greater than 5$\sigma$. Additionally, the zones with high residuals cannot exceed the 
size of the beam. Visual inspection was performed on all maps to confirm. 

\begin{figure}
\begin{center}
\includegraphics[width=.45\textwidth]{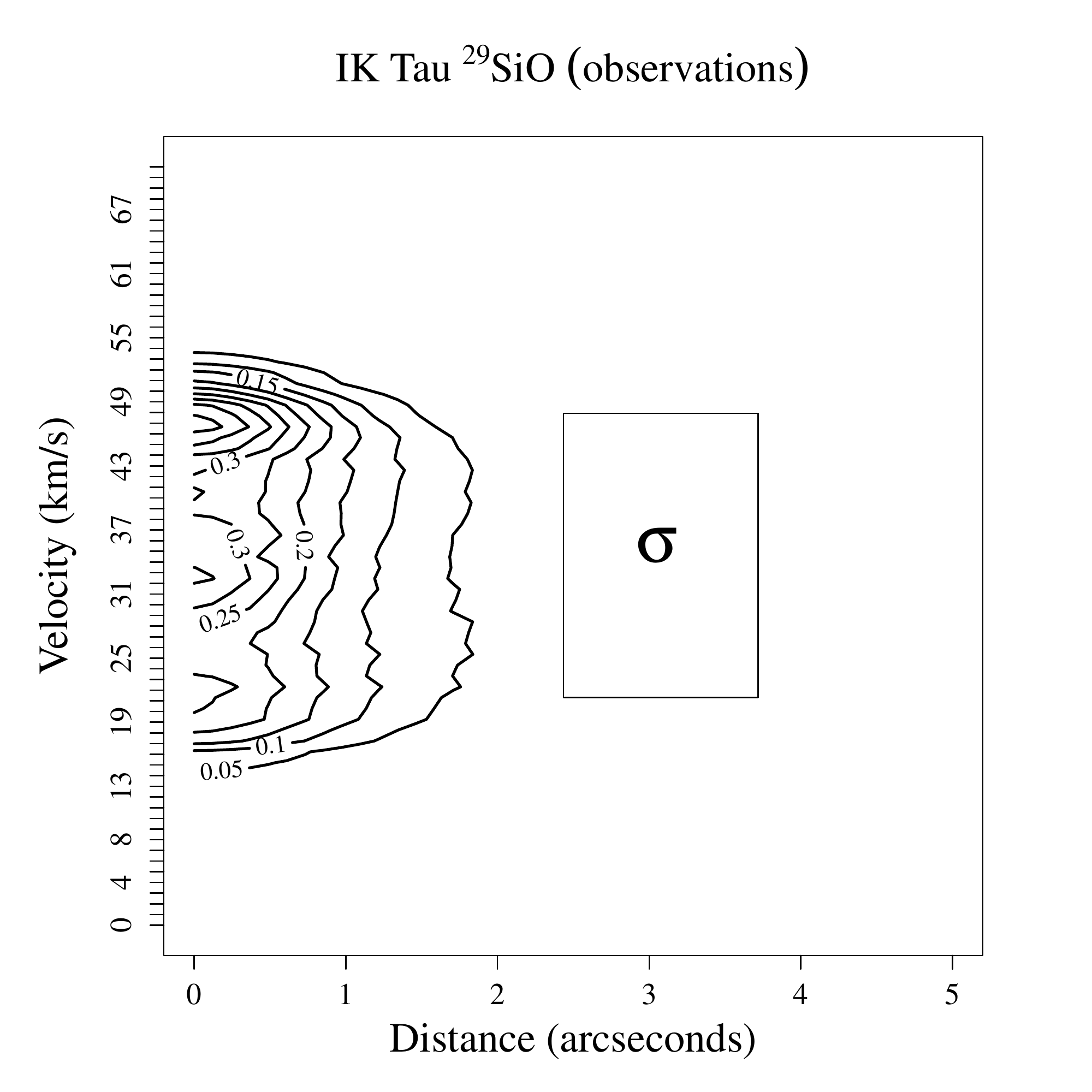}
\end{center}
\caption
{ 
Azimuthally averaged $^{28}$SiO emission as a function of distance and LSR velocity for IK Tau. 
The rectangle outside the contours 
represents the area where the r.m.s. ($\sigma$) was calculated. We chose this region to take into
account errors that could have resulted from the cleaning process.  
}
  \label{fig:rmsreg29}
\end{figure}

\begin{figure}
\begin{center}
\includegraphics[width=.45\textwidth]{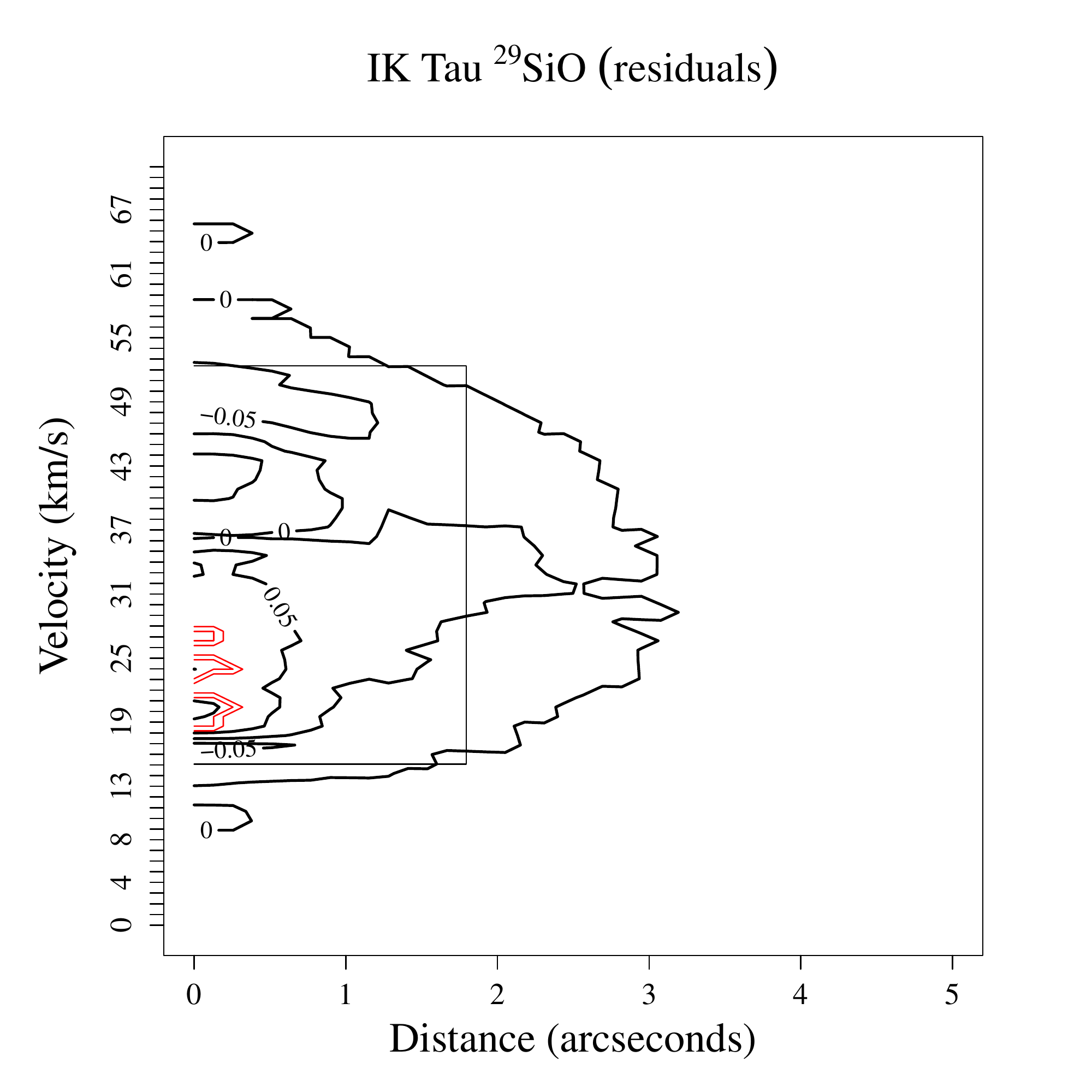}
\end{center}
\caption
{ 
Residual map that results from subtracting the observations from the model, for IK Tau, $^{29}$SiO emission. 
The r.m.s. of the residuals is calculated in the area enclosed by the rectangle.
The red contours illustrate the regions where the maximum residual, in absolute value, is located.  
}
  \label{fig:diff29}
\end{figure}

For the same object, 
but for observations of $^{29}$SiO (See Fig. \ref{fig:rmsreg29}), we calculate $\sigma$
in the region enclosed by a rectangle to the right of the contours. We obtain $\sigma$~$\sim$~0.02~Jy~beam~$^{-1}$, 
smaller than that 
obtained for $^{28}$SiO. When we subtract the model map shown in Fig. \ref{fig:iktaumods} from the observations map,
we obtain the residual map shown in Fig. \ref{fig:diff29}. In this case the r.m.s.in the emission region is 
$\sim$~0.04~Jy~beam$^{-1}$
(2$\sigma$) and the largest value of the residual is $\sim$~0.1~Jy~beam$^{-1}$
(5$\sigma$). Further changing the parameters either 
increases the r.m.s. value of the residual map, or the largest value of the residual for one of the isotopologues, 
therefore we conclude that this is
the model that best fit the observations for both isotopic species.    

The uncertainties for the parameters reported in Table \ref{tab:freeparams} were obtained starting from the previously
established models. We varied slightly one parameter at the time while keeping the others constant. 
Besides visually inspecting the resulting model and residual maps, we calculated the r.m.s. and the maximum
value of the residual map. Once the r.m.s. in the emission region exceeds 3$\sigma$ or 
the maximum value of the residual exceeded 7$\sigma$, we considered that the model is no longer
satisfactory and report the upper and lower uncertainties.      

\begin{figure}[ht]
\begin{center}
\includegraphics[width=.45\textwidth]{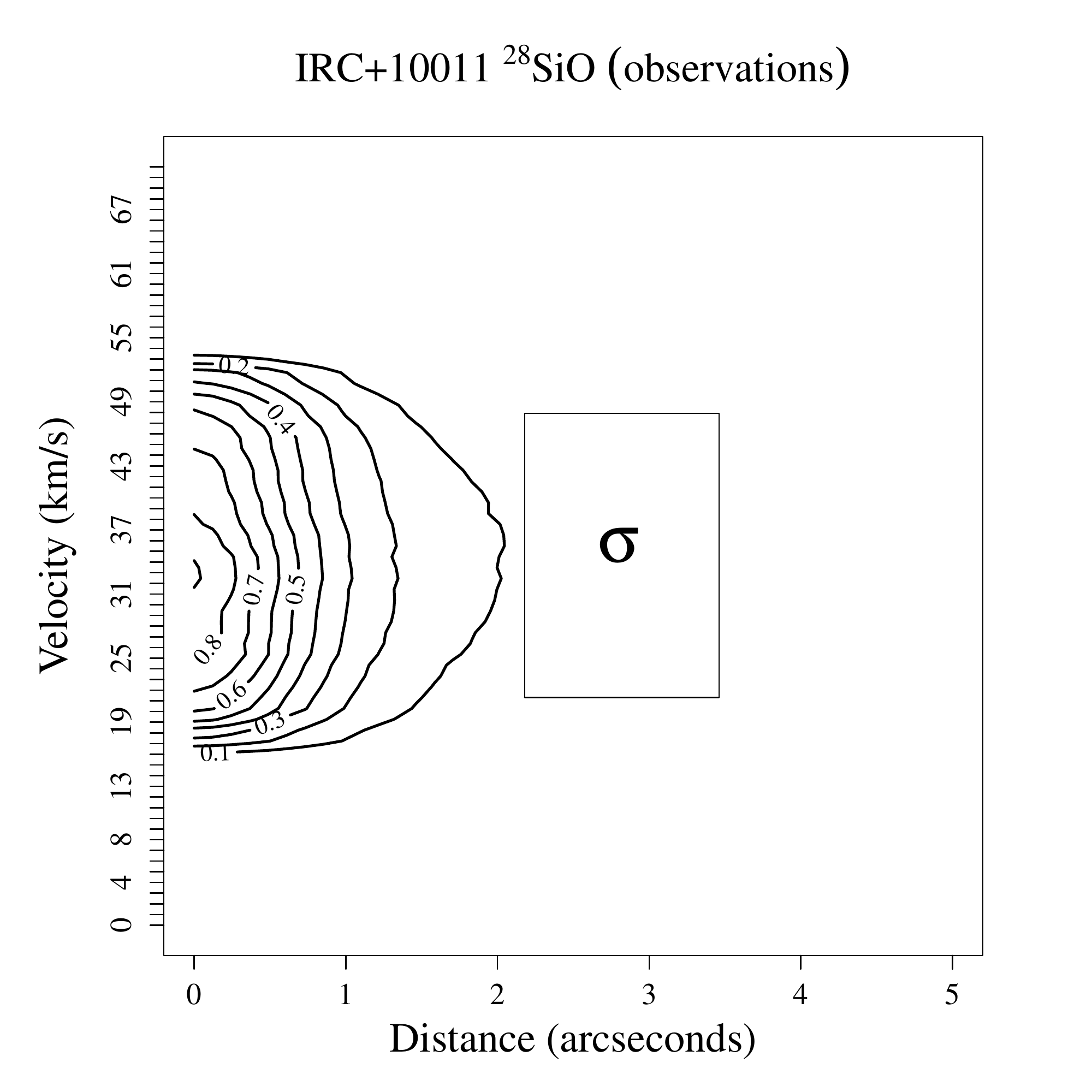}
\end{center}
\caption
{
Azimuthally averaged $^{28}$SiO emission as a function of distance and LSR velocity for IRC+10011.
The rectangle outside the contours
represents the area where the r.m.s. ($\sigma$) was calculated. We chose this region to take into
account errors that could have resulted from the cleaning process.
}
\label{fig:rmsreg28irc}
\end{figure}

\begin{figure}
\begin{center}
\includegraphics[width=.45\textwidth]{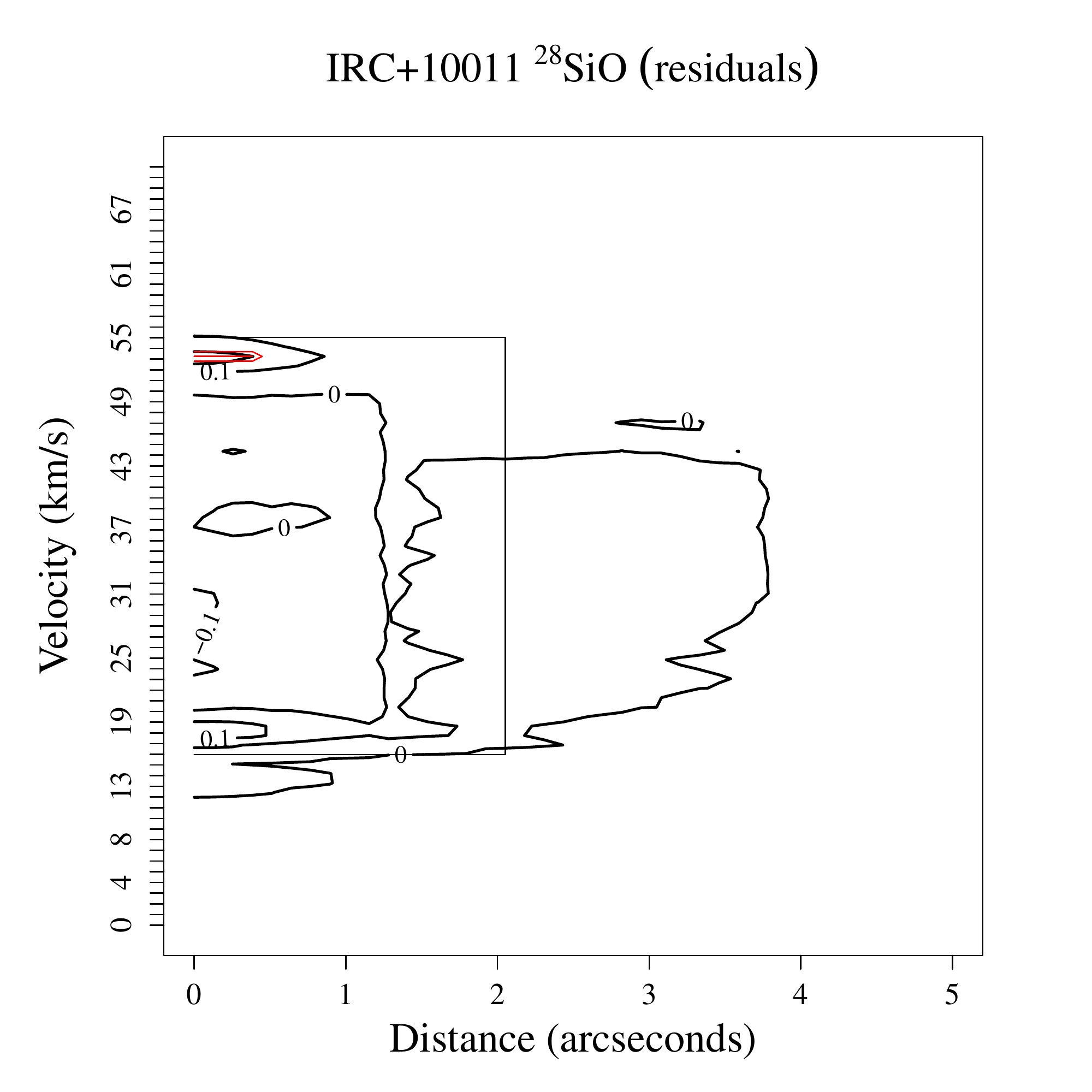}
\end{center}
\caption
{
Residual map that results from subtracting model from observations, for IK Tau, $^{28}$SiO emission. The
r.m.s. of the residuals is calculated in the
area enclosed by the rectangle.
The red contours illustrate the regions where the maximum residual, in absolute value, is located.  
}
\label{fig:diff28irc}
\end{figure}

In the case of IRC+10011, $^{28}$ SiO, the value of the r.m.s. of the observations, calculated in the
rectangular region indicated in Fig. \ref{fig:rmsreg28irc}, is $\sim$ 0.04 Jy beam$^{-1}$; the r.m.s. in the emission
region in the residual map (Fig. \ref{fig:diff28irc}) is 0.05 Jy beam$^{-1}$ (1.25~$\sigma$). The maximum residual value is 
$\sim$~0.2 Jy beam$^{-1}$ (5~$\sigma$).

\begin{figure}[ht]
\begin{center}
\includegraphics[width=.45\textwidth]{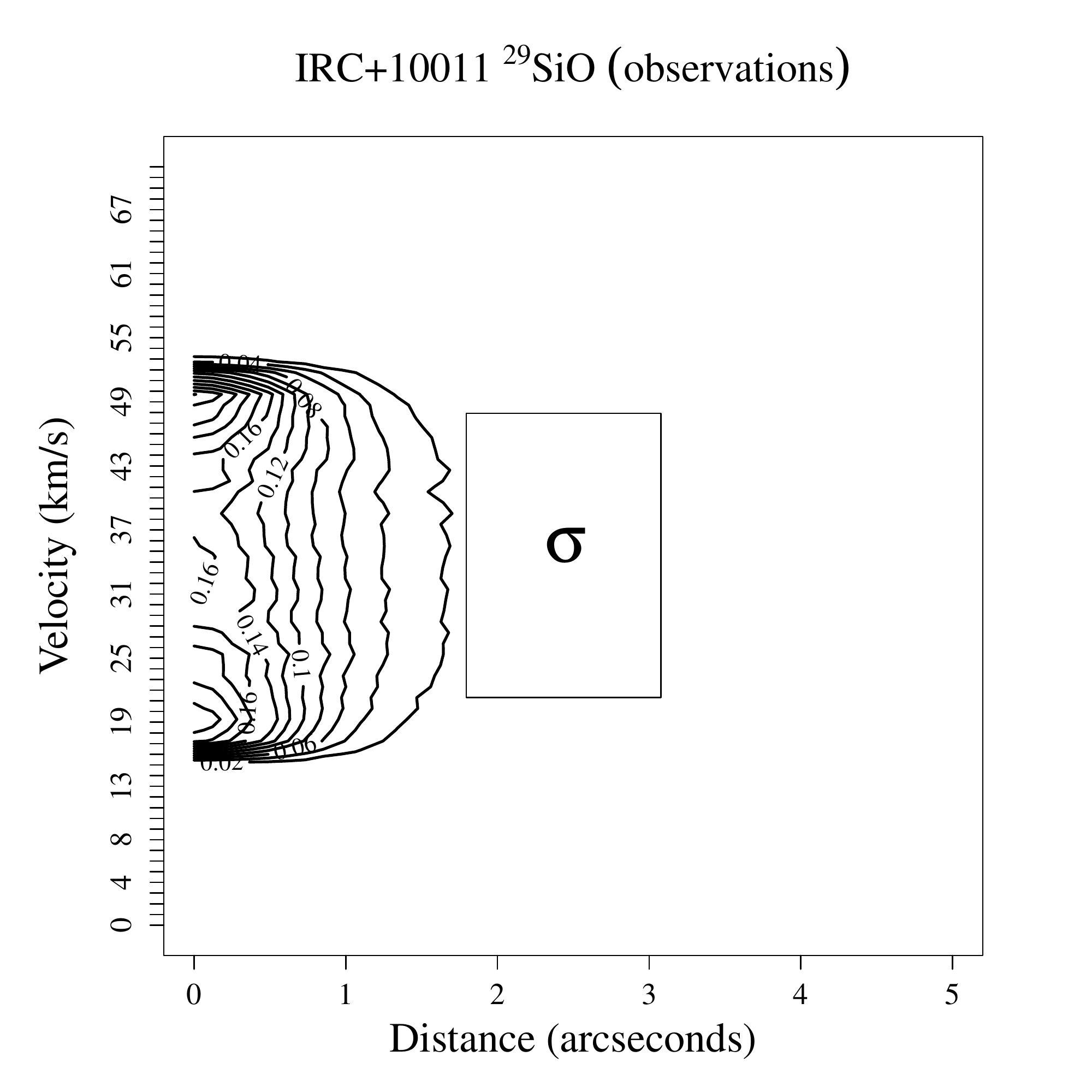}
\end{center}
\caption
{
Azimuthally averaged $^{28}$SiO emission as a function of distance and LSR velocity for IRC+10011.
The rectangle outside the contours
represents the area where the r.m.s. ($\sigma$) was calculated. We chose this region to take into
account errors that could have resulted from the cleaning process.
}
\label{fig:rmsreg29irc}
\end{figure}

\begin{figure}
\begin{center}
\includegraphics[width=.45\textwidth]{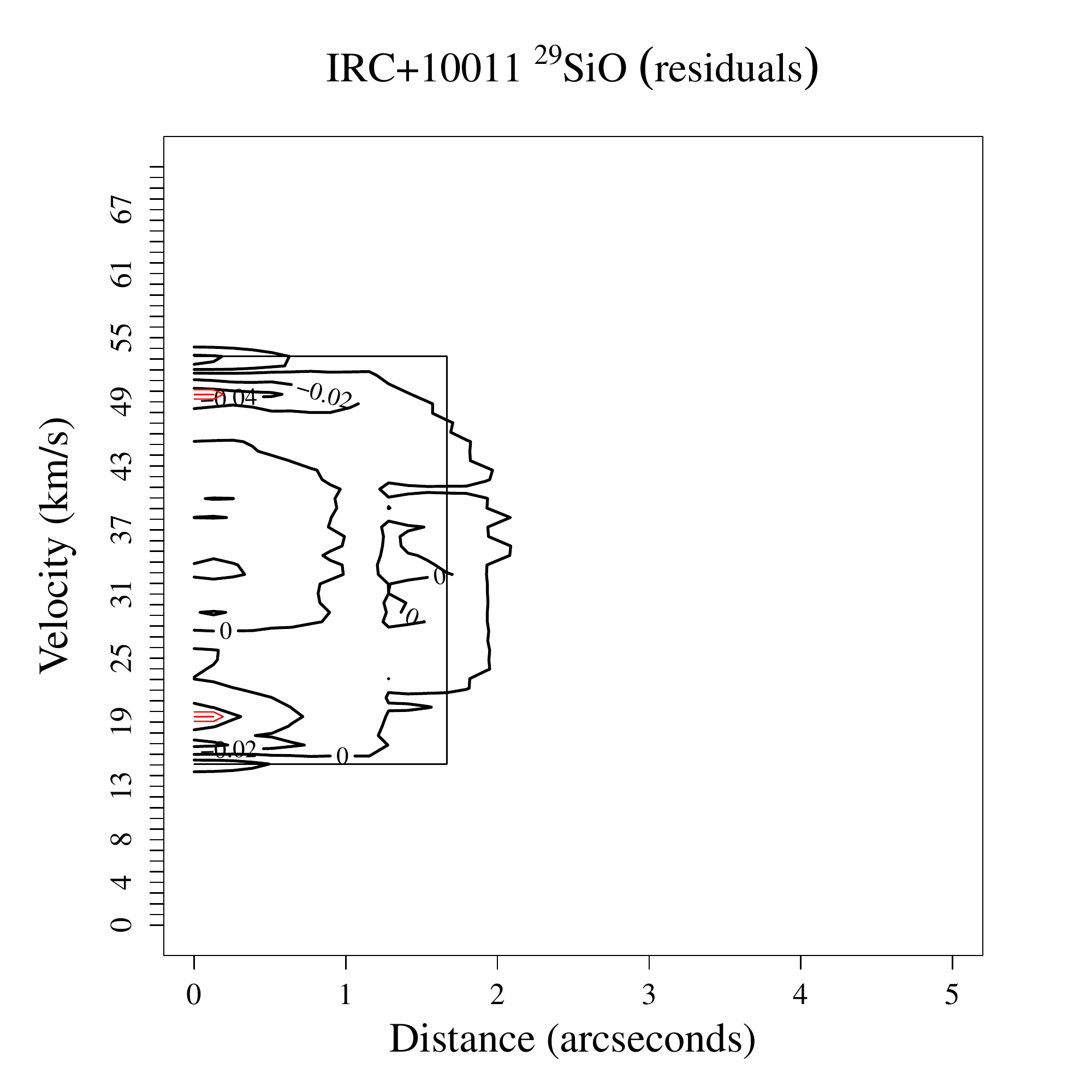}
\end{center}
\caption
{
Residual map that results from subtracting the observations from the model, for IRC+10011, $^{28}$SiO emission. The
r.m.s. of the residuals is calculated in the
area enclosed by the rectangle.
The red contours illustrate the regions in which the maximum residual, in absolute value, is located.  
}
  \label{fig:diff29irc}
\end{figure}

For $^{29}$ SiO emission in the case of IRC+10011, we obtain a r.m.s. value of the observations $\sim$~0.012~Jy~beam$^{-1}$
(See Fig. \ref{fig:rmsreg29irc}). The r.m.s. in the emission region in the residual map (Fig. \ref{fig:diff29irc}) 
is $\sim$~0.015~Jy~beam$^{-1}$
(1.25~$\sigma$). The maximum residual value is $\sim$~0.06~Jy~beam$^{-1}$ (5~$\sigma$).   

Given the simplicity of the models, the complexity of the objects, and the infinite number of
possible parameterisations, in particular of the SiO abundance,
we cannot claim that the resulting models provide a unique fit to the observations. We can only say that
they describe a plausible scenario for the conditions in the inner envelope of M-type AGB stars.

\end{appendix}


\begin{thebibliography}{}


\bibitem[\protect\citeauthoryear{Bose et al.}{2010}]{Bose2010}
Bose, M., Floss, C., \& Stadermann., F. J. 2010, ApJ, 714, 1624 

\bibitem[\protect\citeauthoryear{Bromley et al.}{2014}]{Bromley2014}
Bromley, S. T., Goumans, T. P. M., Herbst, E., Jones, A. P. \& Slater, B. 2014, Phys. Chem. Chem. Phys., 16, 18623 

\bibitem[\protect\citeauthoryear{Bujarrabal \& Alcolea}{1991}]{Bujarrabal1991}
Bujarrabal, V., Alcolea, J. 1991, A\&A, 251, 536 

\bibitem[\protect\citeauthoryear{Bujarrabal et al.}{1989}]{Bujarrabal1989}
Bujarrabal, V., G\'omez-Gonz\'alez, J., \& Planesas, P. 1989, A\&A, 219, 256

\bibitem[\protect\citeauthoryear{Busso et al.}{1999}]{Busso1999}
Busso, M., Gallino, R., \& Wasserburg, G. J. 1999, ARA\&A, 37, 239 

\bibitem[\protect\citeauthoryear{Decin et al.}{2010}]{Decin2010}
Decin, L., De Beck, E., Br\"unken, S., et al. 2010, A\&A, 411, 123

\bibitem[\protect\citeauthoryear{Decin et al.}{2018}]{Decin2018}
Decin, L., Richards, A. M. S., Danilovich, T., Homan, W., \& Nuth, J. A. 2018,
A\&A, 615, A28 

\bibitem[\protect\citeauthoryear{Duari et al.}{1999}]{Duari1999}
Duari, D., Cherchneff, I., \& Willacy, K. 1999, A\&A, 341, L47

\bibitem[\protect\citeauthoryear{Forrest et al.}{1975}]{Forrest1975}
Forrest, W. J., Gillet, F. C., \& Stein, W. A. 1975, ApJ, 195, 423  

\bibitem[\protect\citeauthoryear{Gaia collaboration}{2018}]{Gaia2018}
Gaia Collaboration 2018, ArXiv e-prints, 1804.09365

\bibitem[\protect\citeauthoryear{Gail \& Sedlmayr}{1984}]{Gail1984}
Gail, H. P., \& Sedlmayr, E. 1984, A\&A, 132, 163

\bibitem[\protect\citeauthoryear{Gail et al.}{2016}]{Gail2016}
Gail, H. P., Scholz, M., \& Pucci, A. 2016, A\&A, 591, A17

\bibitem[\protect\citeauthoryear{Gobrecht et al.}{2016}]{Gobrecht2016}
Gobrecht, D., Cherchneff, I., Sarangi, A., Plane, J. M. C., \& Bromley, S. T. 2016, A\&A, 585, A6      

\bibitem[\protect\citeauthoryear{Gonz\'alez-Delgado et al.}{2003}]{Gonzalez2003}
Gonz\'alez-Delgado, D., Olofsson, H., Kerschbaum, F., et al. 
2003, A\&A, 411, 123  

\bibitem[\protect\citeauthoryear{Goumans \& Bromley}{2012}]{Goumans2012}
Goumans, T. P. M., \& Bromley, S. T. 2012, MNRAS, 420, 3344

\bibitem[\protect\citeauthoryear{H\"ofner \& Olofsson}{2018}]{Hofner2018}
H\"ofner, S., \& Olofsson, H. 2018, A\&ARv, 26, 1  

\bibitem[\protect\citeauthoryear{Homan et al.}{2018}]{Homan2018}
Homan, W., Danilovich, T., Decin, L., et al. 2018, A\&A, 614, A113


\bibitem[\protect\citeauthoryear{Karavikova et al.}{2013}]{Karavikova2013}
Karavikova, I., Wittkowsky, M., Ohnaka, K., et al. 2013, A\&A, 560, A75     

\bibitem[\protect\citeauthoryear{Kervella et al.}{2015}]{Kervella2015}
Kervella, P., Montarg\`es, M., Lagadec, E., et al. 2015, A\&A, 578, A77

\bibitem[\protect\citeauthoryear{Khouri et al.}{2015}]{Khouri2015}
Khouri, T., Waters, L. B. F. M., de Koter, A., et al. 2015, A\&A, 577, A114

\bibitem[\protect\citeauthoryear{Lamers \& Cassinelli}{1999}]{Lamers1999}
Lamers, H. J. G. L., \& Cassinelli, J. P. 1999, Introduction to Stellar Winds (Cambridge University Press)  

\bibitem[\protect\citeauthoryear{Li et al.}{2016}]{Li2016}
Li, X., Millar, T. J., Heays, A. N., et al. 2016, A\&A, 588, A4


\bibitem[\protect\citeauthoryear{Lucas et al.}{1992}]{Lucas1992}
Lucas, R., Bujarrabal, V., Guilloteau, S., et al. 1992, A\&A, 262, 491

\bibitem[\protect\citeauthoryear{Molster et al.}{2010}]{Molster2010}
Molster, F. J., Waters, L. B. F. M., \& Kemper, F. 2010 in Lect. Not. Phys. 815, ed. T. Henning 
(Berlin: Springer Verlag), 143   

\bibitem[\protect\citeauthoryear{Monson et al.}{2017}]{Monson2017}
Monson, N. N., Morris, M. R. \& Young, E. D. 2017, ApJ, 839, 123

\bibitem[\protect\citeauthoryear{Morris et al.}{1983}]{Morris1983}
Morris, M., Lucas, R., Omont, A. 1983, A\&A, 142, 107

\bibitem[\protect\citeauthoryear{Nguyen et al.}{2010}]{Nguyen2010}
Nguyen, A. N., Nittler, L. R., Stadermann, F. J., Stroud, R. M., \& Alexander, M. O. 2010, ApJ, 719, 166  


\bibitem[\protect\citeauthoryear{Olofsson et al.}{1982}]{Olofsson1982}
Olofsson, H., Johansson, L. E. B., Hjalmarson, A., Nguyen-Q-Rieu 1982, A\&A, 107, 128 

\bibitem[\protect\citeauthoryear{P\'egouri\'e \& Papoular}{1985}]{Pegourie1985}
P\'egouri\'e, B., \& Papoular, R. 1985, A\&A, 142, 451  

\bibitem[\protect\citeauthoryear{Plane}{2013}]{Plane2013}
Plane, J. M. C. 2013, Roy. Soc. Lond. Phil. Trans. Ser. A., 371, 20335


\bibitem[\protect\citeauthoryear{Sahai \& Bieging}{1993}]{Sahai1993}
Sahai, R., \& Bieging, J. H. 1993, AJ, 105, 595

\bibitem[\protect\citeauthoryear{Sch\"onberg}{1988}]{Schonberg1988}
Sch\"onberg, K. 1988, A\&A, 195, 198

\bibitem[\protect\citeauthoryear{Sch\"oier et al.}{2006}]{Schoier2006}
Sch\"oier, F. L., David, F., Olofsson, H., Zhang, Q., \& Patel N. 2006, ApJ, 649, 2, 965

\bibitem[\protect\citeauthoryear{Sch\"oier et al.}{2004}]{Schoier2004}
Sch\"oier, F. L., Olofsson, H., Wong, T., Lindqvist, M., \& Kerschbaum, F. 2004, A\&A, 422, 651

\bibitem[\protect\citeauthoryear{Sch\"oier et al.}{2005}]{Schoier2005}
Sch\"oier, F. L., van der Tak, F. F. S., van Dishoek, E. F., \& Black, J. H. 2005, A\&A, 432, 369


\bibitem[\protect\citeauthoryear{Velilla-Prieto et al.}{2017}]{Velilla2017}
Velilla-Prieto, L., S\'anchez-Conteras, C., Cernicharo, J., et al. 2017, A\&A, 597, A25  

\bibitem[\protect\citeauthoryear{Vinkovic et al.}{2004}]{Vinkovic2004}
Vinkovi\'c, D., Bl\"ocker, T., Hofmann, K.-H., Elitzur M., \& Weigelt, G. 2004, MNRAS, 352, 3, 852

\bibitem[\protect\citeauthoryear{Vollmer et al.}{2009}]{Vollmer2009}
Vollmer, C., Brenker, F. E., Hoppe, P., \& Stroud, R. M. 2009, ApJ, 700, 774 


\bibitem[\protect\citeauthoryear{Zhao-Geisler et al.}{2012}]{Zhao2012}
Zhao-Geisler, R., Quirrenbach, A., K\"ohler, R., \& Lopez, B. 2012, A\&A, 545, A56

 


\end{thebibliography}
\end{document}